\documentclass[12pt,preprint]{aastex}

\usepackage{natbib}

\usepackage{graphics}
\usepackage{longtable}
\graphicspath{{Figures/}}

\newcommand{\mjbmaxval}{\ensuremath{-18.27\pm0.07}}
\newcommand{\mhbmaxval}{\ensuremath{-18.07\pm0.03}}
\newcommand{\mkbmaxval}{\ensuremath{-18.30\pm0.11}}

\newcommand{\mjbmaxptelval}{\ensuremath{-18.29\pm0.09}}

\newcommand{\mhbmaxptelval}{\ensuremath{-18.08\pm0.04}}

\newcommand{\mkbmaxptelval}{\ensuremath{-18.32\pm0.11}}

\newcommand{\tbmax}  {\ensuremath{t_{B {\rm max}}}}

\newcommand{\OM}     {\ensuremath{{\Omega}_{\rm M}}}
\newcommand{\OL}     {\ensuremath{{\Omega}_{\Lambda}}}
\newcommand{\irbmax}  {\ensuremath{{\rm IR}_{\rm B max}}}

\newcommand{\hbmax}  {\ensuremath{H_{\rm B max}}}

\newcommand{\mhbmax} {\ensuremath{M_{H_{\rm B max}}}}

\newcommand{\jhk}    {\ensuremath{JHK_{s}}}

\newcommand{\sneia}  {\mbox{SNe~Ia}}
\newcommand{\snia}   {\mbox{SN~Ia}}
\newcommand{\chisqnu}{\ensuremath{\chi^2/{\rm DoF}}}
\newcommand{\chisq}  {\ensuremath{\chi^2}}
\newcommand{\kms}    {{}km~s$^{-1}$}
\newcommand{\kmsmpc} {{}km~s$^{-1}$~Mpc$^{-1}$}
\newcommand{\lcdm}   {\ensuremath{\Lambda{\rm CDM}}}

\newcommand{\numsneianew}{18}
\newcommand{\numsneianewall}{21} \newcommand{\numsneialit}{23}
\newcommand{\numsneia}{41}

\newcommand{\numsneiaflow}{28}

\newcommand{\numptelobs}{1087}

\newcommand{\hrms}{0.16}
\newcommand{\hflowrms}{0.15}
\newcommand{\hptelrms}{0.15}

\newcommand{\jrms}{0.29}

\newcommand{\jptelrms}{0.33}

\newcommand{\kptelrms}{0.26}

\newcommand{\krms}{0.29}

\begin{document}

\title{Type Ia Supernovae are Good Standard Candles in the Near Infrared: Evidence from PAIRITEL}
\shorttitle{SNeIa as NIR Standard Candles from PAIRITEL}
\shortauthors{Wood-Vasey et al.}

\author{
{W.~Michael~Wood-Vasey}\altaffilmark{1}, 
{Andrew~S.~Friedman}\altaffilmark{1}, 
{Joshua~S.~Bloom}\altaffilmark{2,3}, 
{Malcolm~Hicken}\altaffilmark{1}, 
{Maryam~Modjaz}\altaffilmark{2},
{Robert~P.~Kirshner}\altaffilmark{1}, 
{Dan~L.~Starr}\altaffilmark{2}, 
{Cullen~H.~Blake}\altaffilmark{1}, 
{Emilio~E.~Falco}\altaffilmark{1}, 
{Andrew~H.~Szentgyorgyi}\altaffilmark{1}, 
{Peter~Challis}\altaffilmark{1}, 
{St{\'e}phane~Blondin}\altaffilmark{1},
{Kaisey~S.~Mandel}\altaffilmark{1},
and
{Armin~Rest}\altaffilmark{4,5}
}

\email{wmwood-vasey@cfa.harvard.edu, afriedman@cfa.harvard.edu,
jbloom@astro.berkeley.edu, mhicken@cfa.harvard.edu,
mmodjaz@astro.berkeley.edu, rkirshner@cfa.harvard.edu,
dstarr@astro.berkeley.edu, cblake@cfa.harvard.edu,
efalco@cfa.harvard.edu, saint@cfa0.cfa.harvard.edu,
pchallis@cfa.harvard.edu, sblondin@cfa.harvard.edu,
kmandel@cfa.harvard.edu,
arest@physics.harvard.edu}

\altaffiltext{1}{Harvard-Smithsonian Center for Astrophysics, 
60 Garden Street, Cambridge, MA 02138
}

\altaffiltext{2}{Department of Astronomy, University of California
Berkeley, Berkeley, CA 94720
}

\altaffiltext{3}{Alfred P. Sloan Research Fellow}

\altaffiltext{4}{Cerro Tololo Inter-American Observatory (CTIO), Colina el Pino S/N, La Serena, Chile}
\altaffiltext{5}{Physics Department, Harvard University, 17 Oxford Street, Cambridge, MA 02138
}

\date{\today}

\begin{abstract}
We have obtained \numptelobs{} near-infrared (NIR; \jhk{})
measurements of \numsneianewall{} Type Ia supernovae (\sneia{}) using
PAIRITEL, the 1.3-m Peters Automated InfraRed Imaging TELescope at
Mount Hopkins, Arizona.
This new set of observations nearly doubles the number of well-sampled NIR
\snia{} light curves.
These data strengthen the evidence that \sneia{} are excellent
standard candles in the NIR, 
even without correction for optical light-curve shape.
We construct fiducial NIR templates for normal \sneia{} from our sample, 
excluding only the three known peculiar \sneia{}: SN~2005bl, SN~2005hk, and SN~2005ke.
The $H$-band absolute magnitudes in this sample of
\numsneianew{} \sneia{} have an intrinsic RMS of only \hptelrms{}~mag
with no correction for light-curve shape.
We found a relationship between the $H$-band extinction and
optical color excess of $A_H=0.2 E(B-V)$.
This variation is as small as the scatter in distance modulus
measurements currently used for cosmology that are based on optical
light curves after corrections for light-curve shape 
Combining the homogeneous PAIRITEL measurements with
\numsneialit{}~\sneia{} from the literature, these \numsneia{}~\sneia{}
have standard $H$-band magnitudes with an RMS scatter of \hrms{}~mag.  
The good match of our sample with the literature sample suggests there are few systematic problems with the photometry.
We present a nearby NIR Hubble diagram that shows no correlation of the
residuals from the Hubble line with light-curve properties.
Future samples that account for optical and NIR light-curve shapes,
absorption, spectroscopic variation, or host-galaxy properties may
reveal effective ways to improve the use of \sneia{} as distance
indicators.  Since systematic errors due
to dust absorption in optical bands remain the leading difficulty in
 the cosmological use of supernovae, 
the good behavior of \snia{} NIR light curves and 
their relative insensitivity to reddening 
make these objects attractive 
candidates for future cosmological work.

\end{abstract}


\section{Introduction}

Type Ia supernovae (\sneia{}) observed at optical wavelengths in the rest frame have
played a leading role in extragalactic astronomy and cosmology in the
past decade.  \sneia{} have been the key to
measuring the Hubble constant \citep{freedman01,jha99,riess05}
and demonstrating cosmic acceleration \citep{riess98,perlmutter99}. 
These discoveries rest on a foundation of photometric and spectroscopic similarities
between high- and low-redshift \sneia{}
\citep{hamuy96,riess99,goldhaber01,hook05,blondin06,jha06,conley06,garavini07,garg07,bronder07,foley07,ellis07}.  
The history of cosmic expansion has been traced out to redshift $z \sim 1$
\citep{tonry03,knop03,barris04}.  The predicted turnover to the matter-dominated era
and the corresponding deceleration at $z>1$ have been
seen by space-based work from {\em HST}~\citep{riess07}.  
Two current projects: ESSENCE
(Equation of State: SupErNovae trace Cosmic Expansion;
\citealt{miknaitis07,wood-vasey07}), and SNLS (SuperNova Legacy Survey;
\citealt{astier06}) seek to constrain the nature of the
dark energy responsible for cosmic acceleration by measuring the
equation-of-state parameter $w$ with large, homogeneous samples of \sneia{}.  
ESSENCE and SNLS data are both
consistent with dark energy described by a cosmological constant,
$w=-1.0$ to better than $\pm0.1$.

These cosmological results are based on measurements of optical emission in the supernova rest frame, but recent
work suggests that \sneia{} may be superior distance indicators in
the near infrared (NIR), with a narrow distribution of peak \jhk{} magnitudes and 4--8 times less sensitivity to reddening
\citep{meikle00,krisciunas04a,krisciunas07}.  
Reddening and absorption of supernova
light by dust in their host galaxies poses the single most vexing
systematic question facing \snia{} cosmology (\citealt{wang06,conley07}).  Avoiding these complications would be very desirable, if the properties of the supernovae permit.  These questions can be approached by building a homogeneous and large set of NIR observations.  This paper is a step in that direction.

Early photometric NIR observations
of \sneia{} were made by \citet{kirshner73,elias81,elias85,graham88,frogel87},
with more recent \jhk{} light curves published by
\citet{jha99,hernandez00,valentini03,candia03,krisciunas00,krisciunas01,krisciunas03,benetti04,krisciunas04b,krisciunas04c,krisciunas05a,krisciunas06,elias-rosa06,elias-rosa07,krisciunas07,phillips06,pastorello07a,pastorello07b,prieto07,stritzinger07,wangx08,pignata08,taubenberger08}.
Northern hemisphere work from the CfA Supernova Program
uses observations from the robotic 1.3-m Peters Automated InfraRed Imaging TELescope
(PAIRITEL; \citealt{bloom06b}) at Mount Hopkins, Arizona.  The NIR observations are part of a systematic program of supernova observations that includes dense sampling of supernova spectra \citep{matheson07} and concurrent $UBVri$ optical photometry (\citealt{hicken08}, {\it in prep.}).  This paper is a first report on the NIR data on \sneia{} from PAIRITEL. Similar work is underway in the southern
hemisphere by the Carnegie Supernova Project using observations
at the Las Campanas Observatory in Chile \citep{freedman05, hamuy06}.
The prospect of large homogeneous data sets coupled with progress in modeling \snia{} NIR light curves~\citep{kasen06} raises hopes that \sneia{}, especially in the rest-frame $H$ band, can be developed into the most precise and accurate of cosmological distance probes. 

Recent work by \citet{krisciunas04a} shows that
\sneia{} have a narrow range of luminosity in \jhk{} at
the time of $B$-band maximum light (\tbmax{}) with smaller scatter than
in the $B$ and $V$ bands.  \citet{krisciunas04a} found no correlation between 
optical light-curve shape and intrinsic NIR luminosity.  
The NIR behavior is in sharp contrast with the optical light curves, where a variety of ingenious methods have been devised to reduce the scatter in distance estimates:
the $\Delta m_{15}$ method (\citealt{hamuy96,phillips99,prieto06}), 
the multicolor light-curve shape method (MLCS/MLCS2k2; \citealt{riess96,riess98,jha07}), 
the ``stretch'' method \citep{perlmutter97,goldhaber01}, 
the color-magnitude intercept calibration method (CMAGIC; \citealt{wang03}), 
the spectral adaptive template method (SALT/SALT2; \citealt{guy05,astier06,guy07}),
and SiFTO \citep{conley08}.
For a sample of 16 \sneia{} observed in NIR
bands, \citet{krisciunas04a} found a RMS of $\sigma_J=0.14$,
$\sigma_H=0.18$, and $\sigma_{K_s}=0.12$~mag.  
We set out to construct an independent set of observations to test this remarkable result.

In this paper we present \numsneianewall{}  \sneia{} 
observed with PAIRITEL from 2005--2007.
The \numptelobs{} individual data points in this sample
represent the largest homogeneous set of
\sneia{} to date and double the NIR \snia{} observations in the literature. Data collection with PAIRITEL is
discussed in \S\ref{sec:data}.  Data analysis, including the mosaic
creation process and our photometry pipeline, is discussed in
\S\ref{sec:analysis}.
We construct NIR \jhk{} templates from this new sample as detailed in \S\ref{sec:template} and
fit these templates to the \snia{} light curves 
to derive the \jhk{} magnitudes at the time of $B$-band maximum light.
These magnitudes are remarkably uniform, particularly
in the $J$ and $H$ bands.  Distances are calculated in
\S\ref{sec:redshifts}, and we discuss the standard magnitude of NIR \sneia{}
in \S\ref{sec:standard}.  We then compare our sample to that compiled in \citet{krisciunas04a}
as well as to some more recent \sneia{}~\citep{krisciunas05a,krisciunas06,krisciunas07,phillips06,pastorello07b,stritzinger07,wangx08,elias-rosa06,elias-rosa07,pignata08,taubenberger08}
in \S\ref{sec:litsneia} and show that the $J$ and $H$ bands at maximum light, 
even from this heterogeneous sample, have a small range in absolute magnitude. Our conclusions are summarized in \S\ref{sec:conc}.

\section{Data}
\label{sec:data}

We present \numptelobs{} near-infrared (NIR) measurements with a signal-to-noise ratio $>3$ of \numsneianewall{} nearby \sneia{} obtained from 2005 to 2007 using the f/13.5 1.3-m PAIRITEL at  the F.L. Whipple Observatory on Mount Hopkins, AZ. (Table~\ref{tab:jhkdata}).  
Dedicated in October 2004, PAIRITEL uses the northern telescope of the Two Micron All Sky Survey (2MASS;
\citealt{skrutskie06}) together with the 2MASS southern camera. PAIRITEL is a fully automated, robotic telescope with the sequence of
observations controlled by an intelligent queue-scheduling database \citep{bloom06b}. 
Two dichroics allow simultaneous observing in 
\jhk{} (1.2, 1.6, and 2.2~$\mu$m; \citealt{cohen03}) 
with three 256$\times$256 pixel HgCdTe NICMOS3 arrays.  
The image scale of 2\arcsec/pixel provides a field of view of 
8.53\arcmin$\times$8.53\arcmin\ for each filter. 
Since the supernova observations are conducted with the instrument that 
defines the 2MASS photometric system, 
we use the 2MASS point source catalog \citep{cutri03} to establish 
the photometric zeropoints. Images are obtained with standard 
double-correlated reads with the long (7.8~sec) minus short (51~msec) 
frames in each filter treated as the ``raw'' frame in the reduction pipeline. 
The telescope is dithered ($< 2$\arcmin) every fourth exposure to aid with 
reductions.

 The CfA Supernova Program used PAIRITEL to follow up SNe discovered by optical searches at $\delta
\gtrsim -30$ degrees with $V \lesssim 18$ mag.  These are objects
for which we routinely obtain extensive data sets of optical spectra and CCD photometry using other instruments at Mount Hopkins.  With $\sim30\%$ of the
time on this dedicated robotic telescope available for supernova observations, we observe $\sim3$--$5$ SNe per
night, and obtain host galaxy template reference images for each supernova.  
The first published PAIRITEL supernova observations were of the 
unusual core-collapse SN~2005bf~\citep{tominaga05}.  
There have been previous papers that present the data processing from PAIRITEL~\citep{bloom06a,butler06,yost06,bloom07,blake07,eisner07,modjaz08},
but since NIR photometry remains more challenging than optical photometry
we present a brief sketch of the data analysis path in the next section.

\section{Data Analysis}
\label{sec:analysis}

Data were processed with one of two mosaicking
pipelines.  For fields dominated by large galaxies we used a pipeline
(see \citealt{modjaz07,kocevski07}) that makes use of a bank of existing sky
frames.  For all other fields, we used the standard
pipeline (see \S\ref{sec:mosaics}).  Both pipelines
process the $100$--$150$ images, each of $7.8$~seconds duration, by flat correction, dark and sky subtraction, registration, and stacking to create a final mosaicked image in each of
the \jhk{} filters.  Mosaicked images typically comprise 3 images at each dither position,
with a mosaicked FOV of $\sim$12\arcmin$\times$12\arcmin.  
The raw images are SWarped~\citep{bertin05} using 0.1\arcsec/pixel sub-sampling 
into final mosaics with 1\arcsec/pixel sampling.
Typical 1800-second observations (including slew overhead) 
reach magnitude limits of $\sim18,17.5,17$ mag for
$J$, $H$, and $K_{s}$ respectively.

\subsection{Mosaics}
\label{sec:mosaics}
 
The PAIRITEL camera has no shutter, so dark current cannot be
measured independently, and ``skark'' background frames include both sky and dark
photons. PAIRITEL supernova observations did not include
pointings that alternate between the source and the sky, so skark frames were created for each mosaic.

For large host galaxies with angular size $\gtrsim 2\arcmin$ (in
the 12\arcmin\ FOV), host galaxy contamination prevents a reliable calculation of the skark background  calculated
from a pixel-by-pixel median through the stack of dithered images. Here skark frames are estimated
using a median match to an archive of 
skark frames in relatively empty fields observed on the same night.
The archival skark image is selected as close in time to the supernova
observations as possible, with a median matched in a  
50$\times$50 pixel box, known to have stable pixel properties, 
in the lower left quadrant of the array. 
After the dark plus sky frame is subtracted, images are mosaicked using
the drizzle technique \citep{fruchter02}.

For well-isolated supernovae with host galaxies of angular size
$\lesssim 2$\arcmin (in a 12\arcmin~FOV), skark frames can be
constructed from the science images themselves.  This construction is done by
applying a cubic B-spline spline fit to a complete time series of sky values in each
pixel for that observation.  Since the fitting characteristics of the
spline curve vary by filter, elevation, and weather, parameters are
adjusted to best fit each observation and filter.
The pixels with sources are masked out using object masks 
generated from Source Extractor catalogs extracted from the raw frames.
The pipeline replaces each masked pixel's flux value with
a more simple ``median skark'' pixel value that is the sum
of the median flux value of each frame and a median over time
of the pixel's deviation from this median frame value.

The spline fit parameters define a function that tracks the time
variation of the skark background in each of the 256$\times$256 pixels over
the duration of the dither pattern.  We construct a skark image
at the time of each 7.8-second science exposure by calculating the 
skark value in each pixel at this time and
subtract that skark image from the science image.
For cases where the galaxy occupies a small fraction of the field, 
this process improves the sky subtraction. 
This standard pipeline works better than simply taking the
pixel-by-pixel median through the dithered (unregistered) stack of
science images because it samples the sky variation on time scales that are
shorter than the total exposure time. Mosaicked images are constructed
using SWarp~\citep{bertin05}.
We use SWarp with a simple BILINEAR sampling because such
a kernel is most suited to our under-sampled raw images.

\begin{figure}
\plotone{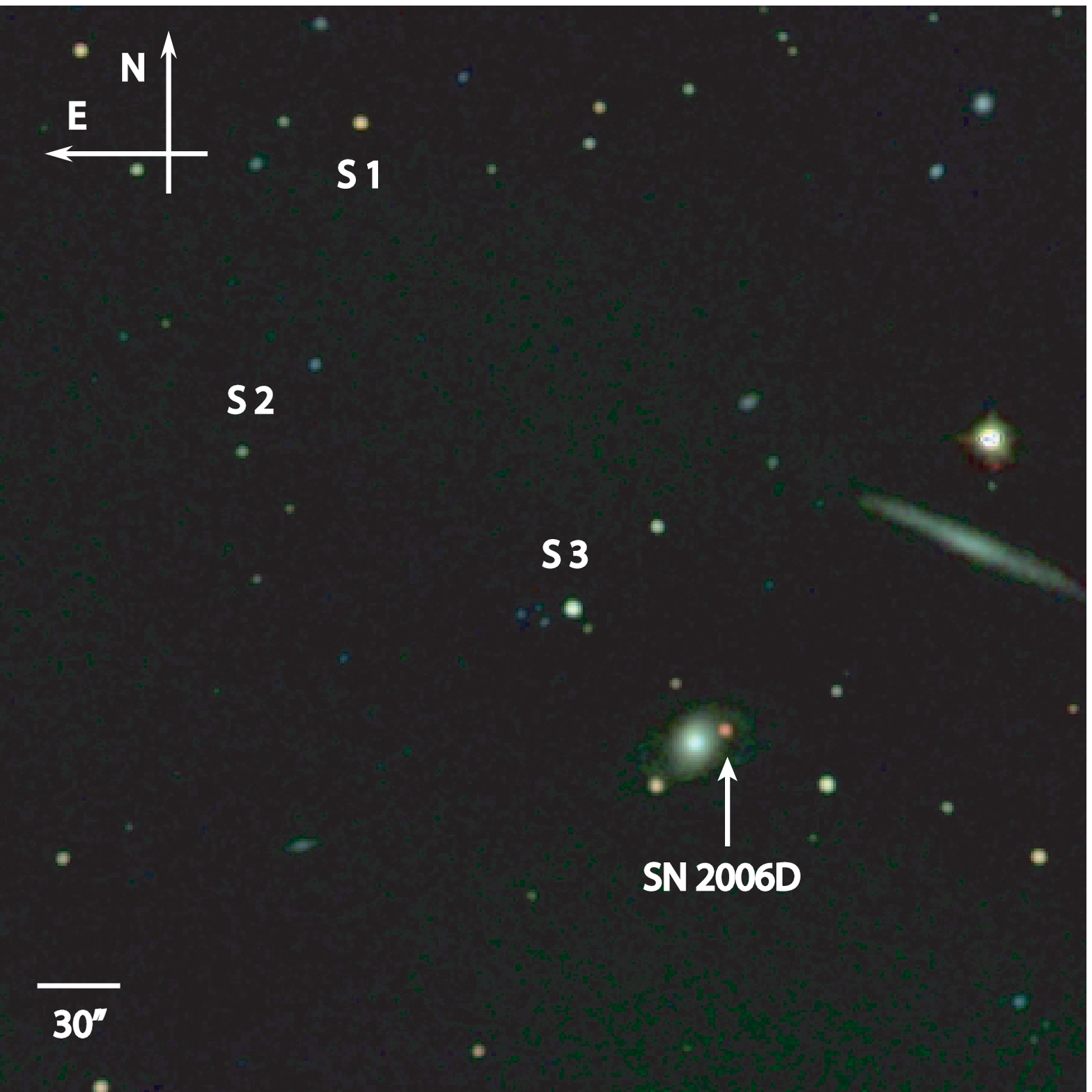}
\caption{PAIRITEL \jhk{} composite color image of SN 2006D 
at a week past maximum light when the \snia{} had NIR magnitudes of 
$(J,H,K)\approxeq(16.1,14.8,15.0)$.  
The image shown is 7.5\arcmin$\times$7.5\arcmin\ in size.
There are $\sim30$ 2MASS stars in this field that were used for the photometric calibration.  The S1, S2, and S3 labels indicate three representative 2MASS stars with \jhk{} magnitudes of S1: (14.11, 13.76, 13.71), S2: (16.02, 15.335, 14.830), and S3 (13.23,  12.61, 12.41).
}
\label{fig:SN2006D}
\end{figure}

Most of our objects were reduced with the standard processing code.  
Only SN~2005cf, SN~2005ke, SN~2005na, and SN~2006X required 
the large galaxy approach.  
For both methods, bad pixel masks
and flat fields were created from archival images.  
Fig.~\ref{fig:SN2006D} shows a final \jhk{} color
mosaic created in the standard processing mode.

\subsection{Photometry}
\label{sec:phot}

Mosaicked images were fed to the photometry pipeline we have used in the
ESSENCE and SuperMACHO projects~\citep{rest05,garg07,miknaitis07}.
This pipeline determines the photon noise for each pixel based on the sky noise in the
mosaic images, registers the images for a supernova to a
common reference frame, and performs point-spread function (PSF) photometry using
DoPHOT~\citep{schechter93}.  The 2MASS catalog~\citep{cutri03} is the natural astrometric and photometric reference system for these
observations.
In a typical 12\arcmin$\times$12\arcmin~field of view,
there were 10--100 2MASS stars in
each filter.  These stars were sufficient to
calibrate the images to the 2MASS system \jhk{} firmly enough that the
underlying uncertainty in the 2MASS system, of about 3\%, is the dominant error
in the photometric calibrations for the light curves presented here.
\citet{cohen03}
describe the 2MASS \jhk{} filter system in detail and \citet{leggett06}
provide color transformations to put observations with the 2MASS
filter system on other widely-used photometric systems.

Final reference images were taken for each galaxy after the supernova had faded.  
DoPHOT
photometry on the images was used for those \sneia{} that were
clearly separated from their host galaxy and had little underlying
contaminating light 
(SN~2005ao, SN~2005cf, SN~2005el, SN~2005hk, SN~2005ke, SN~2005eu, 
SN~2005iq, SN~2005na, SN~2006N, and SN~2006X).  
The standard background annulus subtraction used by DoPHOT 
to estimate the local sky was sufficient to remove any 
remaining contaminating galaxy light.
In every case, the contribution 
from underlying galaxy light was less than 10\% of the \snia{} light at maximum.
Seeing at PAIRITEL is limited by the dome seeing and remains relatively constant
from 1.8--2.0\arcsec.  The good effect of this mediocre dome seeing is that the contaminating galaxy light within the PSF is very nearly constant.

We used subtraction-based photometry (following \citet{miknaitis07}) for 
\sneia{} that were not clearly separated from their host galaxy.
 The NN2 method of
\citet{barris05} was employed (as used in \citealt{miknaitis07}) by 
subtracting all $N\times(N-1)/2$ unique pairs of images to
minimize the sensitivity to subtraction errors.
Because our PAIRITEL data are not critically sampled, the reliability of the PAIRITEL image subtraction is not as good as the subtractions in 
\citet{miknaitis07}.  
Our photometric
pipeline automatically detected and screened out subtractions with significant residual flux in known stars, leaving us with fewer high-quality light-curve points
than would have been obtained in the case of perfect subtractions.
For the supernovae where the underlying galaxy contribution was
a small fraction of the supernova light, direct photometry on the
unsubtracted images proved preferable. 

Photometry was extracted from either the unsubtracted or the subtracted
images by forcing DoPHOT to measure the PSF-weighted flux of the
object at a fixed position.  This position was determined in the $J$-band difference
images that had a signal-to-noise ratio
$>5$.  This average position was
used for fixed-position DoPHOT photometry of each image of a \snia{}.

Flux measurements were calibrated to the 2MASS system by using photometric solutions to the 2MASS catalogs~\citep{cutri03}.  For the
difference images the calibrated zeropoint from the template was used,
with suitable correction for the convolution of the template image as detailed by
\citet{miknaitis07}.
Our NIR PAIRITEL light curves for the \numsneianewall{} \sneia{} presented in this paper are given in Table~\ref{tab:irlcs}.

\section{NIR Template}
\label{sec:template}

We constructed \jhk{} templates with NIR light curves from PAIRITEL,
using our own optical CCD observations \citep{hicken08} to establish the time of B-band maximum, \tbmax{}.

\begin{figure}
\plotone{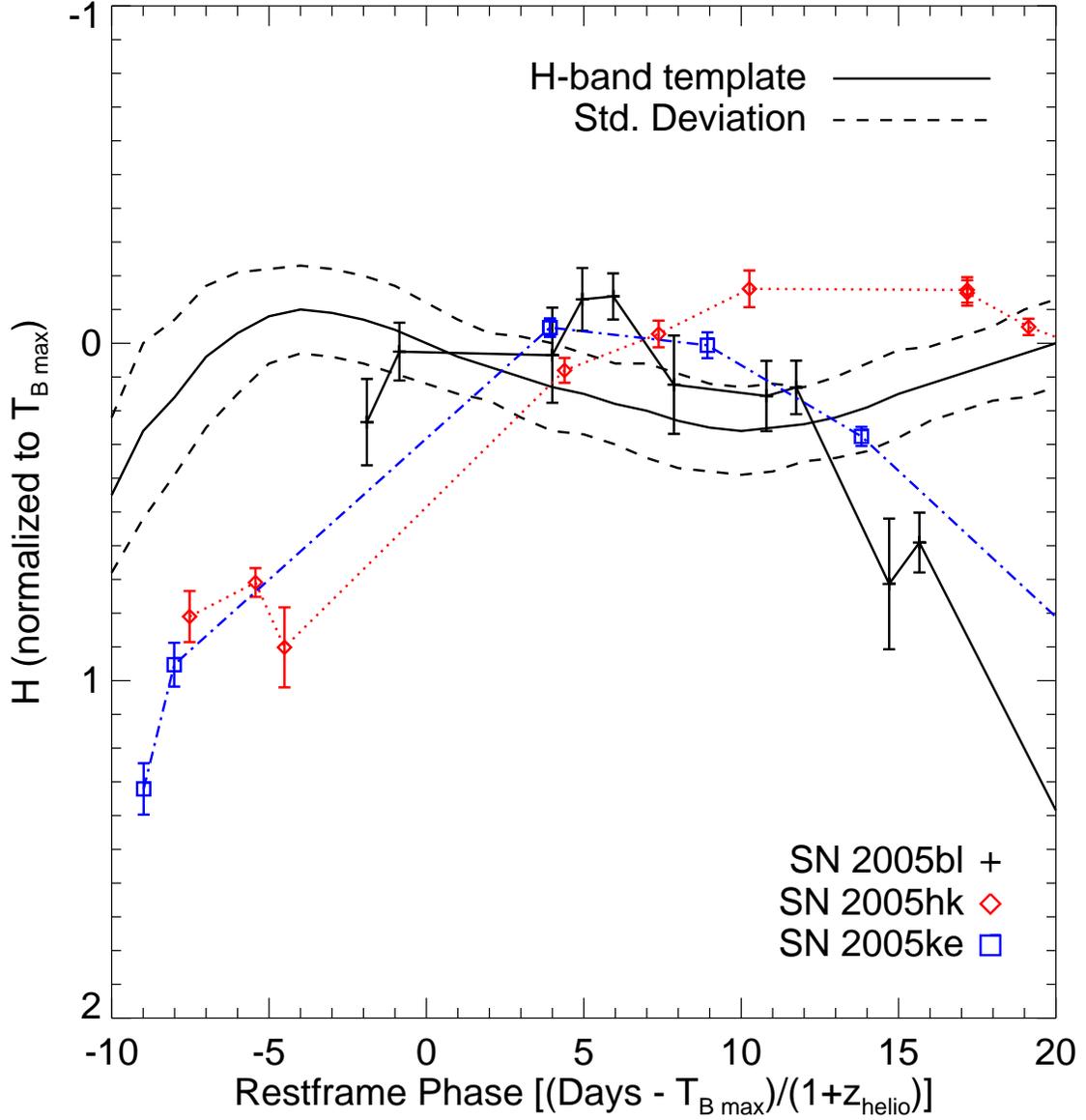}
\caption{SN~2005bl, SN~2005hk, and SN~2005ke were observed as part of the PAIRITEL
campaign but were excluded from the construction of the template
because they were known to be unusual \sneia{}.  
The H-band template we have constructed for this paper is valid 
from -10 to +20 rest-frame days from B-band maximum light.
We compare the $H$-band light curves of these unusual \sneia{} with the normal \snia{} $H$-band template to
demonstrate the clarity with which these unusual supernovae can be distinguished from the
normal \sneia{} used in the Hubble diagram.  
}
\label{fig:weirdsneia_vs_template}
\end{figure}

In constructing the templates, we excluded the three fast-declining \sneia{} from our overall sample of \numsneianewall{} \sneia: 
SN~2005bl, SN~2005hk, and SN~2005ke.  
SN~2005bl has an optical fast decline rate~\citep{taubenberger08} similar to the 
archetypal fast-decliner, SN~1991bg~\citep{filippenko92}.
SN~2005hk is known to be an unusual
\snia{}~\citep{phillips07,sahu07}.
SN~2005ke is a 1991bg-like \snia{}
with possible circumstellar interaction~\citep{patat05,immler06}.
SN~2005bl, SN~2005hk, and SN~2005ke exhibit only 
one infrared hump instead of the two exhibited by normal \sneia{}.
The \jhk{} maximum for these ``dromedary'' light curves
occurs after $B$-band maximum light, unlike normal ``bactrian'' \sneia{}
for which the first NIR maximum occurs $3$--$5$ rest-frame days before $B$-band maximum light.
We compare our \snia{} $H$-band template to SN~2005bl, SN~2005ke, and SN~2005hk in
Fig.~\ref{fig:weirdsneia_vs_template}.  
Because we have excluded these three
objects, our template does not extend to
1991bg-like \sneia{} or other unusual \sneia{}.  
This seems a reasonable approach while the data set is small.  
Unusual supernovae can be identified from a NIR light curve alone, or
they can be identified from their optical spectra and light curves which we obtain as a matter of course.
In any case, we defer incorporating these unusual objects into a more comprehensive treatment of NIR light curves until the
available database of near-infrared \snia{} light curves is more fully populated.

We applied the $K_s$-corrections of \citet{krisciunas04b} to the NIR light curves based on the heliocentric redshift.
The NIR light curves for the remaining \numsneianew{}~\sneia{} were registered to a common
phase by subtracting \tbmax{} and accounting for time dilation
based on the heliocentric redshift, $z_{\rm helio}$~\citep{leibundgut96,goldhaber01}.
Unlike \citet{krisciunas04a}, we do not
further adjust this phase by the optical light-curve width parameter.
We found that compensating for the light-curve width gave no improvement in the H-band template fits (which is the focus of this paper) or in the
resulting dispersion of absolute magnitudes (see \S\ref{sec:standard}) determined between $-10$ and $+20$ days in phase.
However, the position of the $J$-band second NIR maximum is variable, and may be related to intrinsic luminosity. 
In larger data sets and in more detailed investigations of the $J$ band it will be worth exploring whether a width-correction parameter would produce a more effective template for fitting the second maximum at later times.

Between $[-10,+30]$ days, where we have excellent sampling, 
we constructed our lightcurve templates by an unweighted averaging of the points 
from this initial set of standardized light curves into 1-day bins.
No outlier rejection was used in this process.
The \snia{} light curves were then registered to a common magnitude scale by 
fitting each to this zeroth-iteration template
and determining the maximum magnitude at B-band maximum light (\irbmax{}).
The fitting process is insensitive to the determination of the initial \irbmax{}
because this process is repeated with each iteration and the memory of the first
guess is completely erased after the second iteration.
At late times ($>30$~days), 
we had fewer points from which to construct the template,
but it was clear that the \sneia{} were consistent with a linear fit.
For our templates after $+30$ days we adopted linear templates
based on a fit to \snia{} data at these late times with 
a forced connection to the binned value at 30~days.
The complete template was then smoothed twice
with a boxcar of length $5$~days.  
The smoothing has very little effect near maximum light because of our 
dense sampling at those epochs.  Varying the smoothing length between 
$1$ and $5$ days did not substantially affect the template.

To refine this template, we determined the magnitude value at
\tbmax{} that minimized the \chisq{} of the template versus the data over the span from $[-10,+20]$~days.  
These new magnitude offsets were then used to 
seed the process above where we had previously used zero for the \irbmax values.
The procedure was iterated twice more to construct final values at \tbmax{}
for each \snia{}, and the final template was constructed
based on these values.
We found that three iterations were sufficient to reach convergence.
The uncertainty in the template is the standard
deviation of the residuals of the \snia{} light-curve points around the
mean template in a moving 5-day window.

We also explored using a simple linear interpolation on 1-day
sampling and found that the binned template and the linearly-interpolated
templates agreed to within hundredths of a magnitude in this
region near peak.  We take this as a strong indication that
we have reached limit of intrinsic variability of \sneia{} in
the construction of this template since the effective use of
5 times more data in the binned approach yielded the same template
as the linear interpolation, which only uses $\sim20$\% of the available data.

This template construction procedure was performed for each of the \jhk{} NIR passbands.
For our data, the $H$-band light curve was observed to exhibit small scatter from
$[-10,+30]$~days with a particularly tight distribution from $[-10,+20]$~days.  
The $J$-band aggregate light curve has small scatter about the template 
from $[-10,+10]$~days but begins to show variations in the time and flux of the secondary maximum for individual supernovae.
The $K_s$-band data from our PAIRITEL observations is not as good as the
$J$- and $H$-band data due to increased sky background in $K_s$.
Our template in the $K_s$ band shows significantly more scatter with
fewer objects.  
We draw no firm conclusions in this paper with regard to $K_s$-band
\snia{} light curves.

The unusual cases that were excluded from the template construction are easily seen to be far from the H-band template. Fitting SN~2005bl, SN~2005hk, or SN~2005ke with the PAIRITEL \snia{} $H$-band template
results in a sufficiently bad \chisqnu{} ($>3$) (see Table~\ref{tab:hmax}) that these \sneia{} would be excluded
based on that criterion alone.

\section{Galaxy Redshifts}
\label{sec:redshifts}

Recession velocities for our \sneia{} are those of the identified host
galaxies in the NASA/IPAC Extragalactic Database (NED)\footnote{\url{http://nedwww.ipac.caltech.edu/}}.  We use the
quoted velocity with respect to the Virgo infall model of \citet{mould00}.
For \sneia{} within 3000~\kms{} we used available information on distances
to the galaxies to minimize our sensitivity to the details of the local
flows.  
We used the values of \citet{krisciunas04a,krisciunas04b} for 
SN~1998bu and SN~2002bo and \citet{wangx08} for SN~2006X.
After accounting for these \sneia{}, we were left with three \sneia{} within 3000~\kms{} for which we used the \citet{mould00}
values, SN~2003du, SN~2005cf, and SN~2006D.  
Outside of 3000~\kms{} differences among various flow models are not significant.
One supernova, SN~1999cl,
occurred in a Virgo cluster member galaxy, NGC~4501, so we used the Virgo cluster redshift instead of the recession
velocity of NGC~4501.  
We assumed a peculiar velocity of $\sigma_{\rm vel}=150$~\kms{}~\citep{radburn-smith04} and
the redshift measurement uncertainty for each host galaxy as reported by NED.
These uncertainties were converted into a distance modulus uncertainty,
$\sigma_\mu ({\rm vel})$, 
using Eq.~\ref{eq:dztodmu}.
\begin{equation}
\sigma_\mu({\rm vel}) = \frac{5}{z \ln{10}} \sqrt{({\sigma_{\rm vel}/c})^2 + {\sigma_z}^2}
\label{eq:dztodmu}
\end{equation}
This is valid for a smooth cosmology that resembles any
currently allowed cosmological expansion model.  Values for $\sigma_\mu({\rm vel})$ for the \sneia{} considered in this paper are given in column 12 of Table~\ref{tab:hmax}.

We confirmed the NED heliocentric redshift values with the
heliocentric galaxy redshifts derived from spectra of the hosts
as observed by the CfA Supernova Program\footnote{\url{http://www.cfa.harvard.edu/supernova/RecentSN.html}} and measured using the galaxy redshift code RVSAO~\citep{kurtz98}.
All were consistent with the NED values within 200~\kms{}.  For SN~2007cq, no NED redshift was available for
the host galaxy so we used the value from the CfA galaxy spectrum as
the heliocentric velocity and then used the \citet{mould00} model
conversion provided by NED to express this redshift in the CMB+Virgo
infall frame.

\section{Distance Moduli and the Standard NIR Brightness of \sneia}
\label{sec:standard}

The most uniform behavior of our \snia{} light curves is in the $H$-band, and that band is where we focus our analysis.
In the most simplistic treatment, we take the \hbmax{} for each \snia{} to
have a standard value.  Then the distance modulus is simply:
\begin{equation}
\mu = \hbmax - \mhbmax 
\label{eq:observed_distance_modulus}
\end{equation}
where $\mhbmax$ is the absolute $H$-band magnitude of a \snia{} at $B$-band maximum light.
The numerical value for the absolute magnitude of \sneia{} in $H$ is separate
from the question of how well the $H$-band magnitudes trace cosmic expansion.

When we combined the results for the PAIRITEL \sneia{} with our analysis of the literature \sneia{} in \S\ref{sec:litsneia} 
we found that an $H$-band extinction correction of $A_H=0.19 E(B-V)$ best minimized
the correlation between Hubble-diagram residual and $E(B-V)$ as measured by MLCS2k2
and applied that correction to the apparent magnitudes computed from fitting to our $H$-band template.
While, the trend was not formally statistically significant because the 
most highly extinguished \sneia{} are also the ones that are at the lowest
recession velocity and have the most uncertain distances
it was consistent with direct comparison of extinction for \sneia{} in the same galaxy.
Two of the \sneia{} in our sample, SN~2002bo and SN~2002cv,
were both in NGC~3190 and had significant extinction.
When we assumed that each of these \sneia{} had the same intrinsic
brightness, we found the same relationship between $H$-band extinction
and optical color excess of $A_H=0.2 * E(B-V)$.
While quantitatively uncertainty, this $R_H\sim0.2$ corresponds to an equivalent $R_V=1.2$,
which is in qualitative agreement with previous work on observed supernova color-luminosity behavior~\citep[c.f.,][]{perlmutter99,knop03,guy05,conley07,guy07}.
Resolving this mysterious behavior of supernova colors, host galaxy dust, and apparent luminosity is currently the major challenge in \snia{} cosmology.

The set of redshifts, apparent magnitudes, and accompanying
uncertainties from \S\ref{sec:redshifts} 
were compared to a \lcdm{} concordance cosmology 
(\OM{}, \OL{}, $w$)=$(0.27, 0.73, -1)$.  
We solve for the absolute mean brightness of a \snia{} in H-band at B-band maximum light by taking the weighted mean of the differences of the apparent magnitudes and the \lcdm{} distance moduli (the exact cosmology doesn't quantitatively affect the results at these redshifts):
\begin{equation}
\mhbmax = \sum_{i=1}^{N} \frac{\hbmax^{i} - \mu^{i}}{{\sigma^i}_{\hbmax}^2 + {\sigma^i}_\mu({\rm vel})^2} \left / \sum_{i=1}^N \frac{1}{{\sigma^i}_{\hbmax}^2 + {\sigma^i}_\mu({\rm vel})^2} \right.
\label{eq:mhbmax}
\end{equation}
Assuming an $H_o$ of 72~\kmsmpc{}~\citep{freedman01,spergel03,spergel07} 
we find $\mhbmax=\mhbmaxptelval$ with an RMS of \hptelrms{}~mag for the fiducial $H$-band magnitude at B-band maximum light.
A similar analysis for the $J$-band magnitude at 
$B$-band maximum light finds $M_{J_{\rm Bmax}}=\mjbmaxptelval$
with an RMS of \jptelrms{}~mag.  These values are for 
 the new \sneia{} presented here.

For a cosmological fit one would marginalize over the combination
\begin{eqnarray}
\mathcal{M}_{\hbmax} & = & \mhbmax - 5 \log_{10}{h_o} + 15 + 5 \log_{10}{c} \\
\sigma_{\mathcal{M}_{\hbmax}} & = & 5 \frac{\sigma_{h_o}}{h_o} \log_{10} {e} \\  
\label{eq:scriptm}
\end{eqnarray}
where $h = H_o/(100$~\kmsmpc$)$ and $c$ is the speed of light in vacuum in \kms{}.
The marginalization in our case is trivial because we are assuming
a fixed \lcdm{} cosmology.  This assumption allows us to propagate a $H_o$ uncertainty of 10\% directly
through Eq.~\ref{eq:scriptm} to determine our global uncertainty in $\mhbmax$. The mean value of \mhbmax{} is unchanged by the marginalization, but the total uncertainty is increased to $\pm0.22$.  
The uncertainty in $H_o$ dominates the overall uncertainty of $\mhbmax$.

Small variation in the NIR absolute magnitudes of \sneia{} is an important result that was 
first found by \citet{krisciunas04a}. Here, we have an independent confirmation of this result
from a homogeneous set of well-sampled \sneia{} light curves observed with PAIRITEL.  In the next section, we expand that data set by joining the PAIRITEL sample with \sneia{} from the literature.

\section{Comparison with Literature \sneia}
\label{sec:litsneia}

\begin{figure}
\plottwo{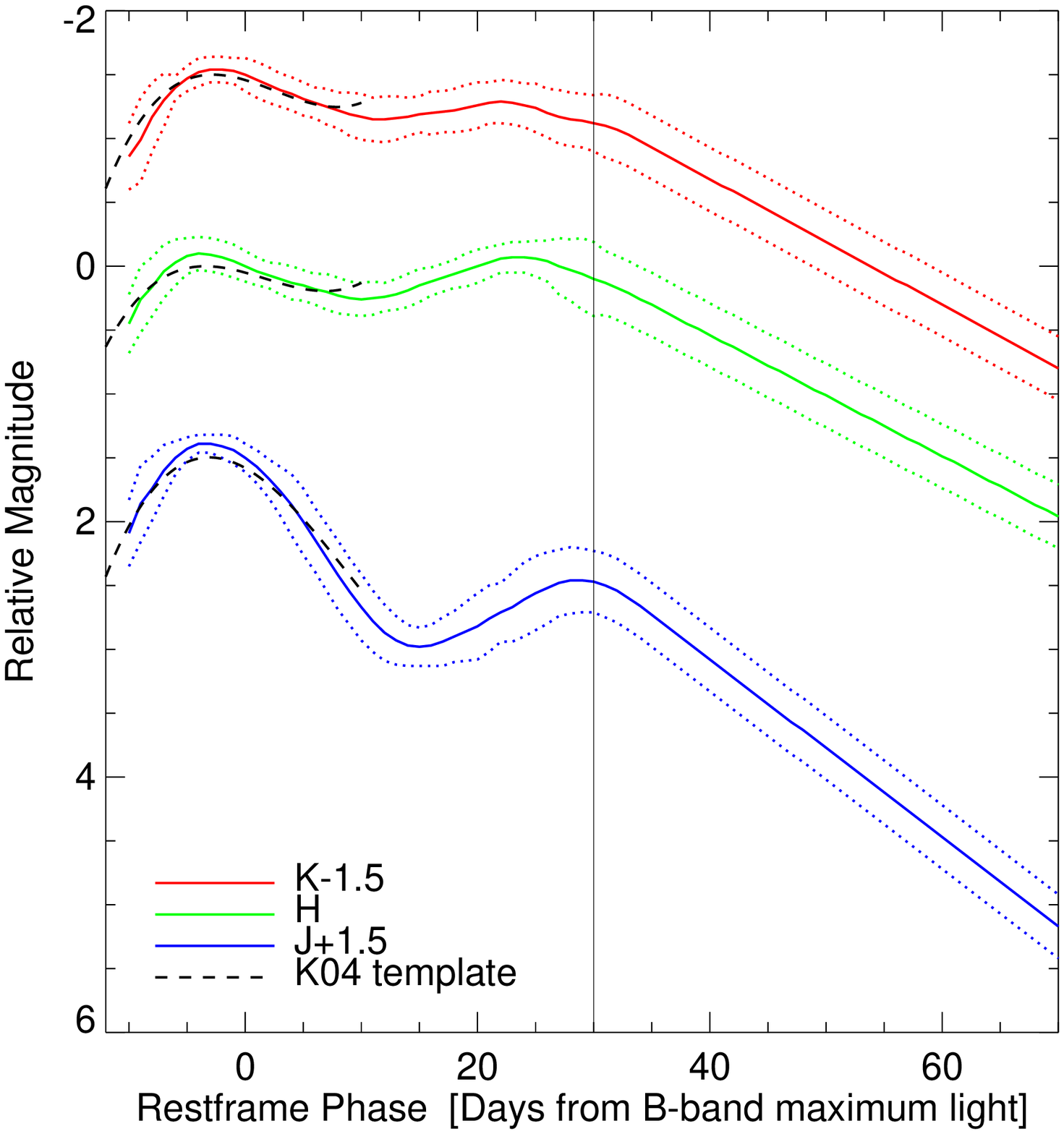}{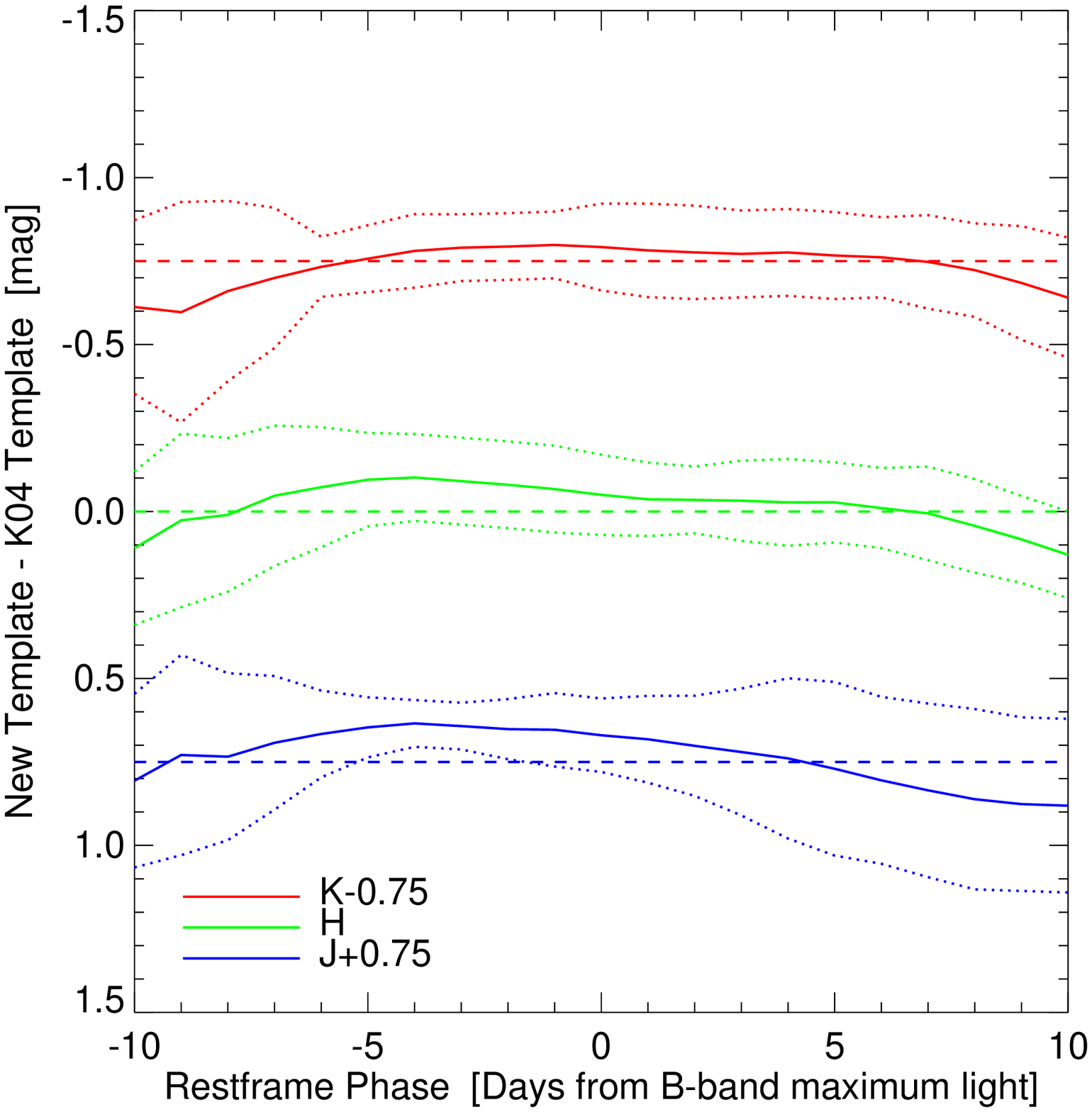}
\caption{\snia{} \jhk{} templates.
The $1$-$\sigma$ uncertainties (dashed lines) are based on the sample
variance within a 5-day moving window of the \sneia{} with respect to the
fiducial template.  The vertical line marks the time after which a linear fit was used to fit the data (0.3~mag uncertainties).  The 3rd-order polynomial templates of \citet{krisciunas04a} are plotted in dashed lines.  The right-hand figure shows the difference between our template and the template of \citet{krisciunas04a}.  While the polynomial templates show approximate agreement with
the new templates presented here, there is a noticeable difference in the
H-band lightcurve before B-band maximum light.  This difference is largely responsible for the different values of H-band maximum in this paper and \citet{krisciunas04a}.
}
\label{fig:template}
\end{figure}

\begin{figure}
\plotone{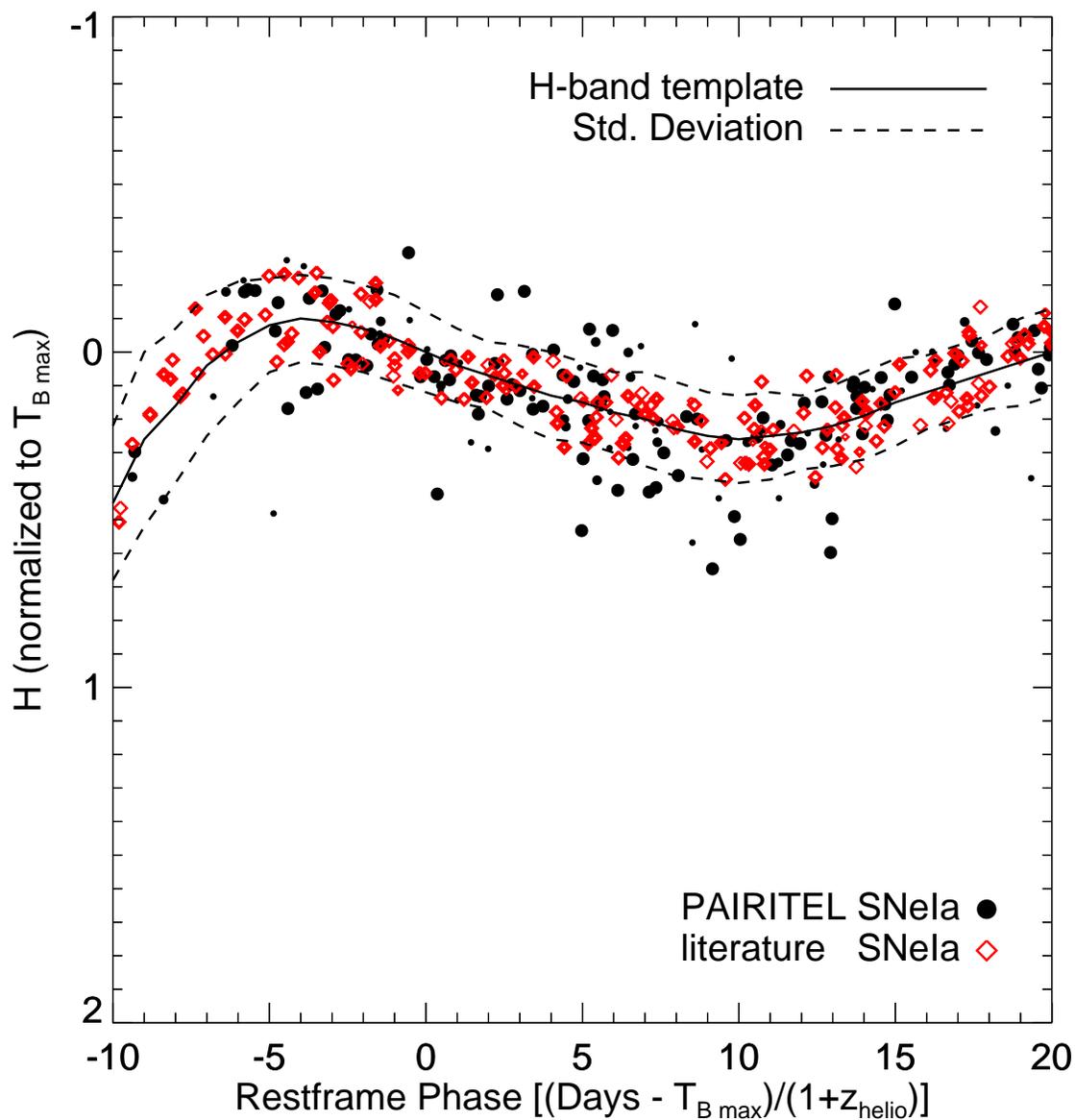}
\caption{NIR $H$-band template based on all of the \sneia{} considered in this paper
(excluding the dromedary \sneia{} shown in Fig.~\ref{fig:weirdsneia_vs_template})
together with the constituent \snia{} observations.  Model uncertainties are
indicated by dashed lines as in Fig.~\ref{fig:template}. 
For clarity data error bars are not displayed.  Instead points are scaled by their
signal-to-noise ratio: small -- $10\ge$SNR$>3$; medium -- $20\ge$SNR$>10$; large -- SNR$>20$.
}
\label{fig:template.H}
\end{figure}

We joined our PAIRITEL sample with \numsneialit{}~\sneia{} from the literature 
(see Table~\ref{tab:hmax}) extending back to SN~1998bu (earlier observations
of supernovae with earlier detectors were not included in this study).
When creating the template including the literature \sneia,
SN~1999ac was excluded because it was classified
as an unusual \snia{}~\citep{phillips06}.
The PAIRITEL \sneia{} listed in \S~\ref{sec:template}
were similarly excluded from the global template-generation.
The literature \sneia{} that were published on the \citet{persson98} system
were converted to the 2MASS system using the conversion equations of
\citet{carpenter01}.  These corrections were typically on the order
of $0.01$~mag.
We generated a new set of NIR templates following the same prescription
as in \S~\ref{sec:template} and found that the full sample of \sneia{} was just as
well-behaved as the PAIRITEL sample.  This is a tribute to the care of the observers and the clear definition of the 2MASS photometric system.  
The NIR \snia{} templates and associated standard deviations for this combined sample are
shown in Fig.~\ref{fig:template} and given in Table~\ref{tab:jhk_templates}.
The small scatter  
of the $H$-band light curves from $[-10,+20]$~days and the $J$-band
light curves from $[-10,+10]$~days was preserved in the heterogeneous
sample.  Fig.~\ref{fig:template.H} shows the tight scatter of the
$H$-band \snia{} light-curve data points around the fiducial template.
Over-plotted in Fig.~\ref{fig:template} are the NIR templates of
\citet{krisciunas04a}.  There is general agreement but noticeable
systematic discrepancy in the H-band template before maximum light.
Together with the difference in definition of quoted $H$-band maximum
(in this paper we refer to $M_H$ at B-band maximum light whereas
\citet{krisciunas04a} quote the observed $H$-band magnitude at H-band maximum light), this disagreement in the $H$-band templates explains the bulk of the $0.3$~mag difference between the quoted $M_H$ values in these respective works.

\begin{figure}
\plotone{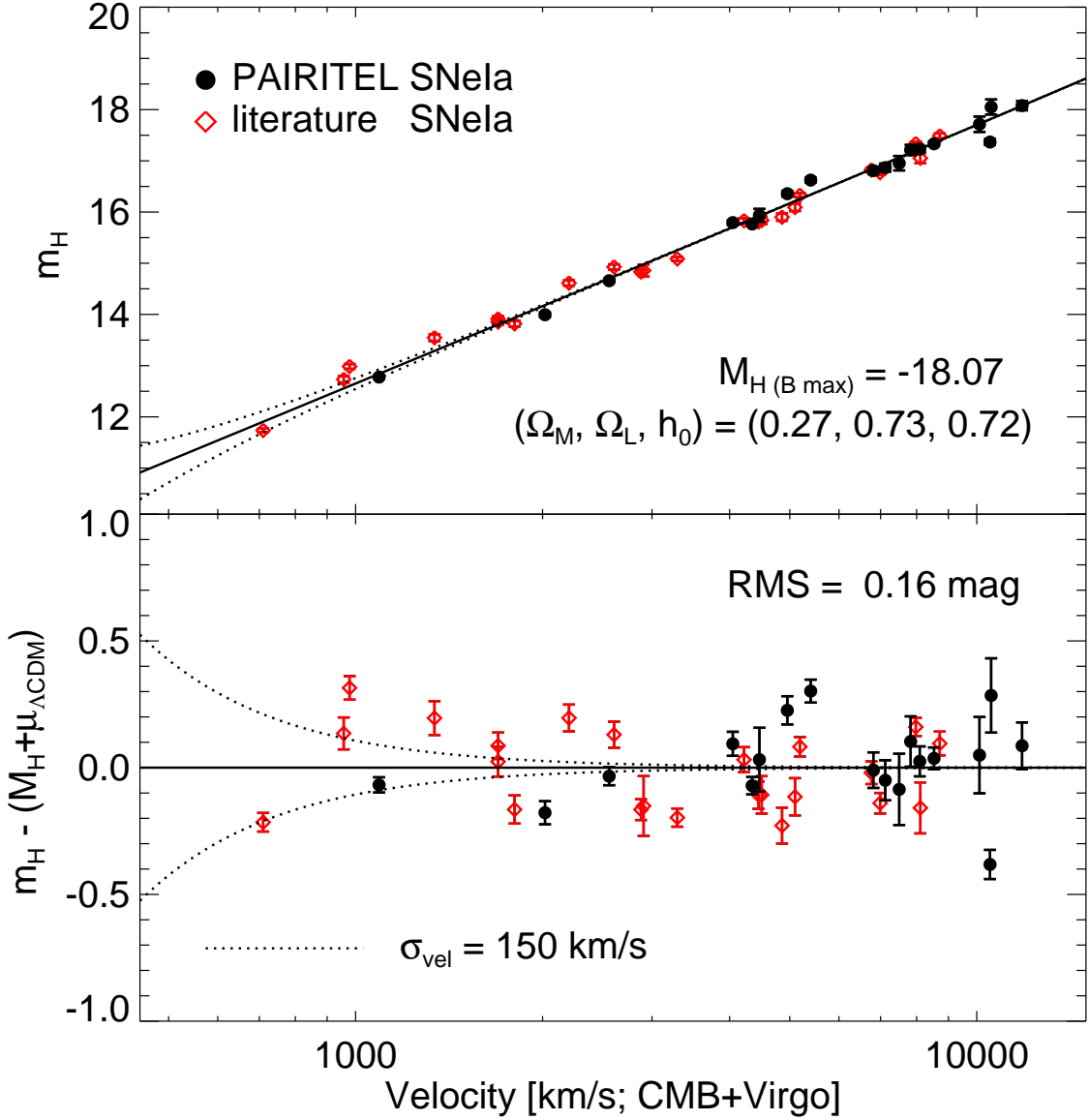}
\caption{$H$-band \snia{} Hubble Diagram ($\mu_H$ vs.\ z and residual vs.\ z).  
RMS dispersion of \hflowrms{}~mag.  
\sneia{} from the literature are shown as open diamonds 
while the new \sneia{} from this paper are shown as filled circles.
The error bars are the fit uncertainties from the fit
to the $H$-band template (column 7 of Table~\ref{tab:hmax}).
The reduced Hubble constant, $h_o$, is $h_o=H_o/100$~\kmsmpc.}
\label{fig:hband_hubble_diagram}
\end{figure}

\begin{figure}
\plotone{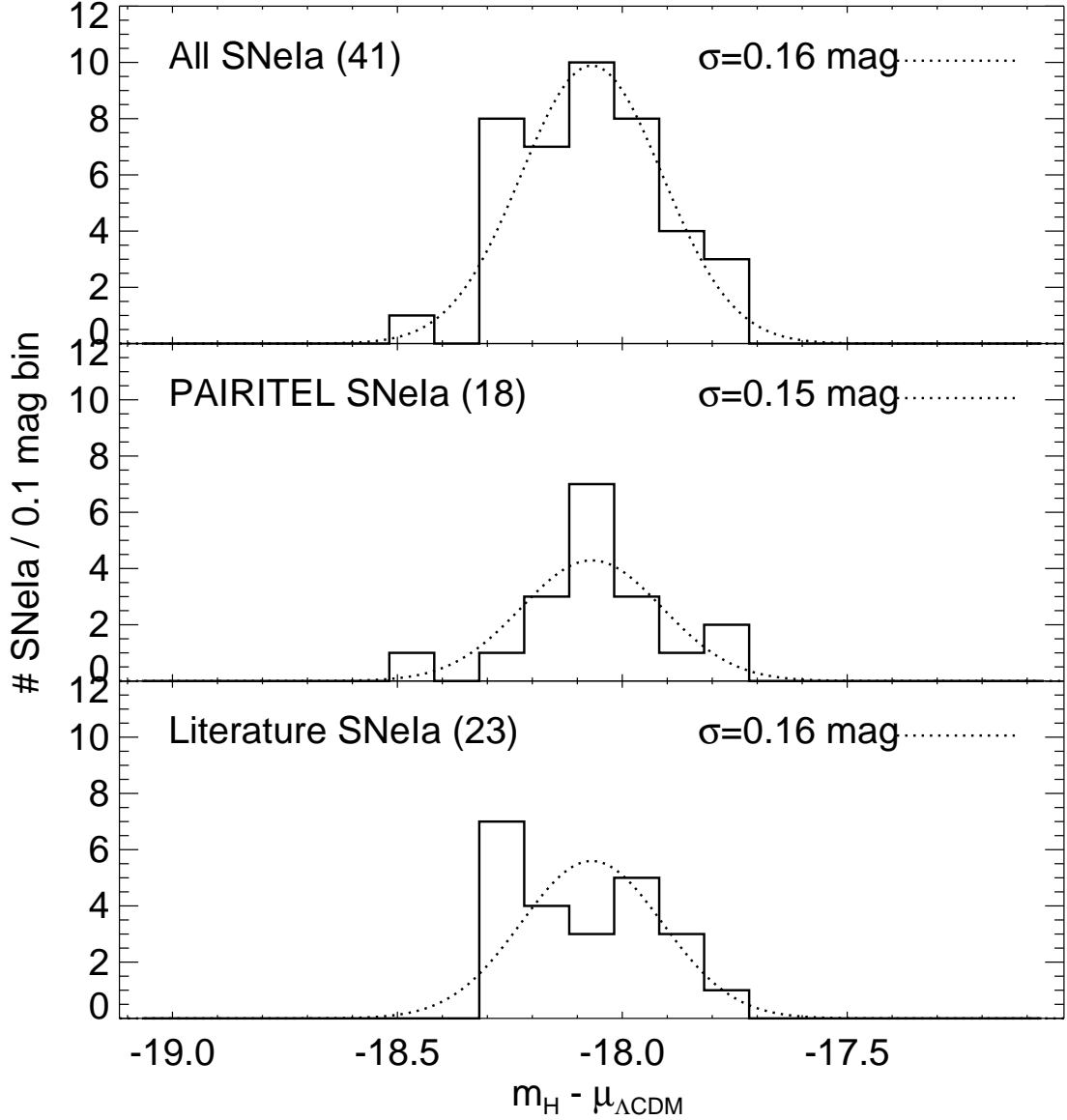}
\caption{A histogram of the derived absolute magnitudes $\mhbmax$ 
based on a \lcdm{} cosmology with $H_o=72$~\kmsmpc for all \sneia{} (top panel)
along with histograms for the PAIRITEL sample (middle panel)
and the literature \sneia{} (bottom panel). 
Overlying each histogram is a Gaussian with the RMS width of the given sample (dotted line)
normalized to the number of \sneia{} in each plot.
}
\label{fig:absolute_mag_hist}
\end{figure}

\begin{figure}
\plotone{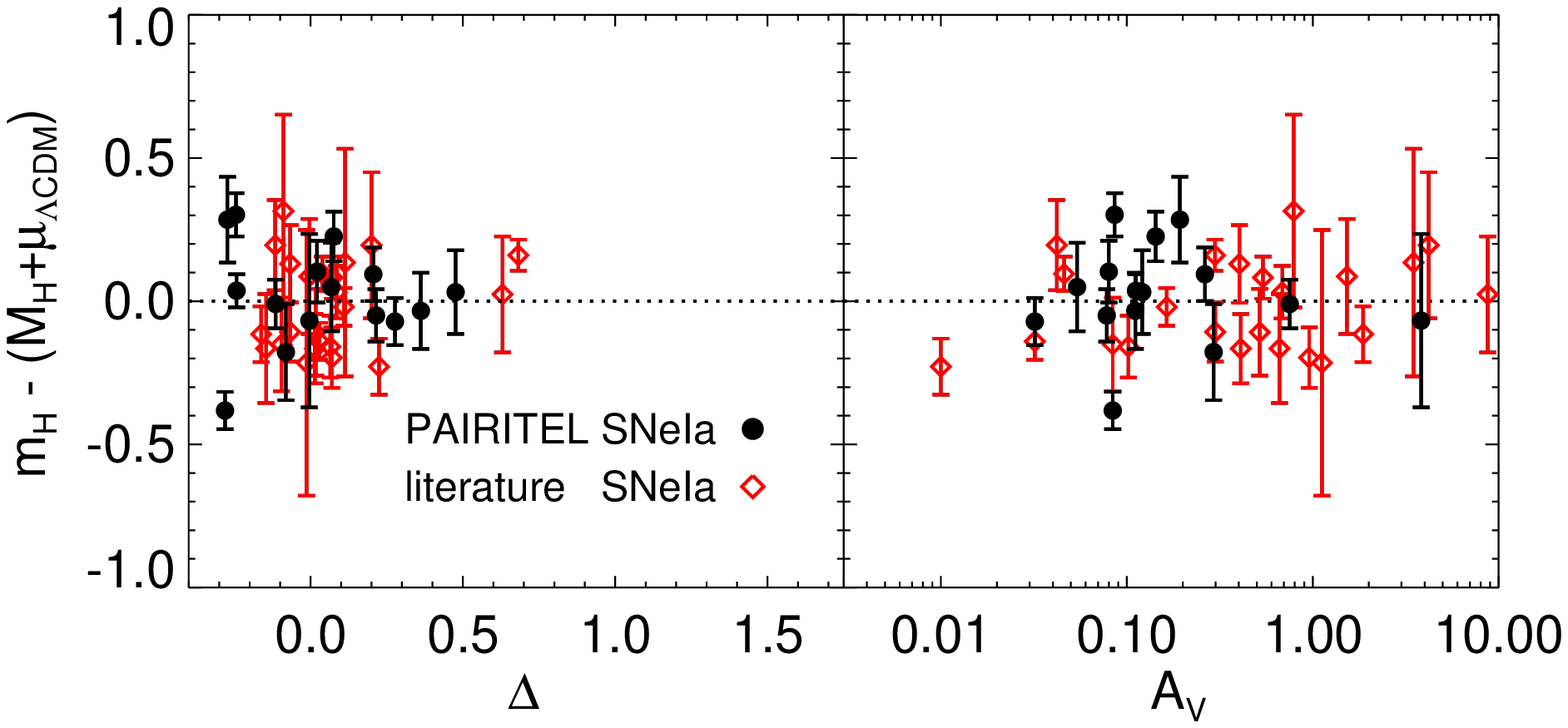}
\plotone{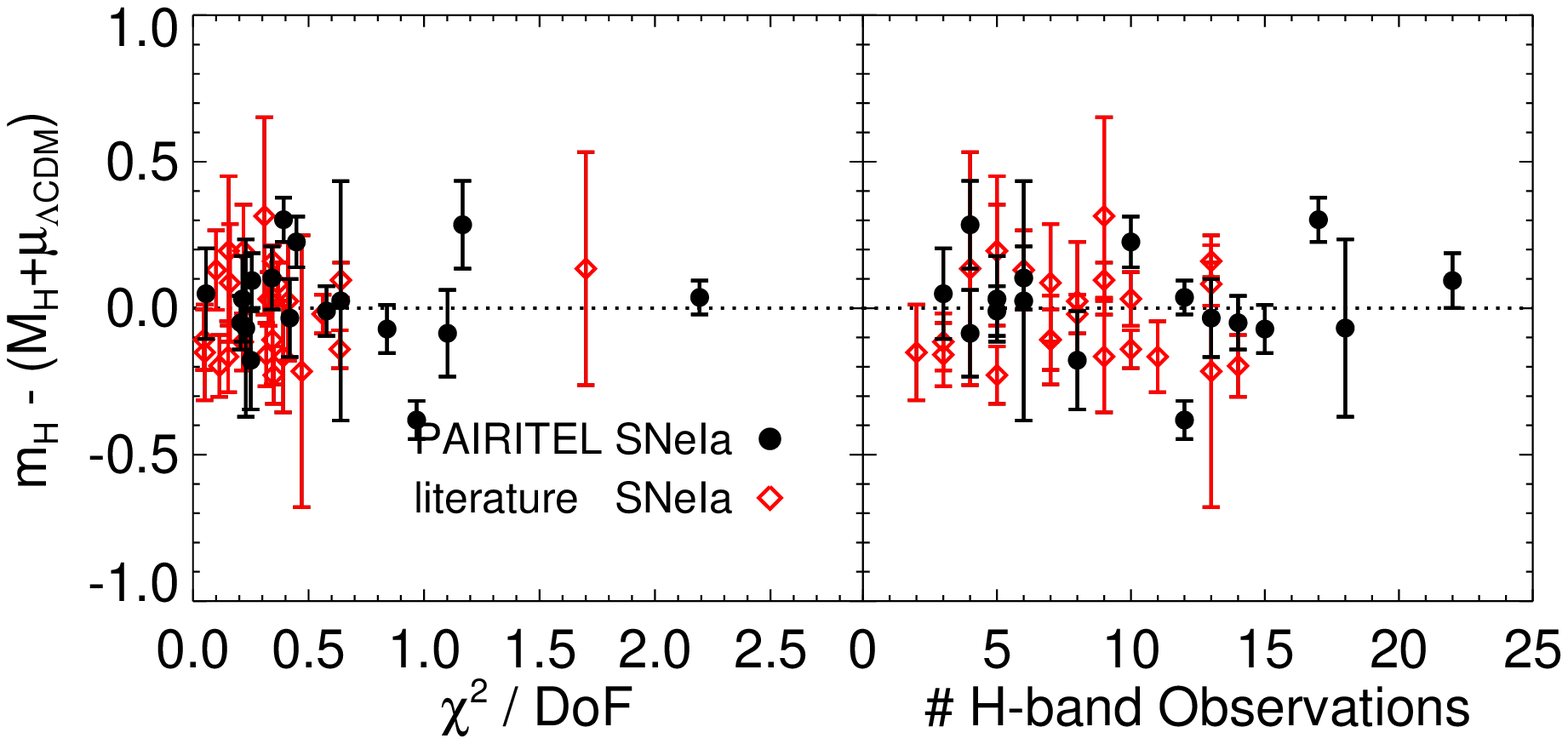}
\caption{Hubble diagram residuals.
(top panel) The Hubble diagram residuals 
of the extinction-corrected $m_H$ as a function of 
the light-curve shape parameter, $\Delta$, 
and the measured optical extinction, $A_V$.
Both plots are are entirely consistent with no
dependence of Hubble diagram residual on $\Delta$
after the extinction correction of $A_H = 0.06 A_V$.
(bottom panel) The Hubble diagram residuals as a function 
of $\chi^2/$DoF 
and the number of H-band light-curve points (DoF$=$\#~light-curve~points$-1$).
These plots demonstrate that the residuals are not sensitive to the 
formal \chisqnu{} of the fit and are reliable even with just a few data points.
The large number of points with \chisqnu$ < 1$ implies that we have likely 
overestimated the uncertainties in our fiducial template or in our photometric observations.
Unlike in Fig.~\ref{fig:hband_hubble_diagram}, 
the error bars shown here represent both the model uncertainties (Table~\ref{tab:hmax}; column 7) 
and the peculiar velocity (150~\kms{}) (Table~\ref{tab:hmax}; column 4) 
added in quadrature (Table~\ref{tab:hmax}, column 12) 
to better allow for a relative comparison of the deviation.
There is a mild correlation between
the size of the error bars and $A_V$ because the \sneia{} with large $A_V$ 
also tend to be nearby ($z<2000$~\kms{}) due to selection effects
against finding highly-extinguished supernovae at larger distances.
}
\label{fig:res_mu_vs_chisq_dof}
\label{fig:res_mu_vs_delta_av}
\end{figure}

We present the $H$-band apparent magnitudes from this global template
in Table~\ref{tab:hmax}.   The comparison of these apparent magnitudes
with the expected luminosity distance relation of a \lcdm{} cosmology,
(\OM{}, \OL{}, $h_o, $w$)=(0.23, 0.77, 0.72, -1)$, 
is shown in Fig.~\ref{fig:hband_hubble_diagram}.  
We find a global $H$-band RMS of \hrms{}~mag around an absolute
magnitude of $M_H=\mhbmaxval$ (see
Fig.~\ref{fig:absolute_mag_hist}).
A similar analysis for $J$-band finds similar agreement with 
the PAIRITEL-only values with a global value of 
$M_J=\mjbmaxval$ with an RMS of \jrms{}~mag.
The $K_s$-band data from PAIRITEL is more limited and only adds 9 
normal \sneia{} to the literature.
The updated global values including these new \sneia{} against
the global template as generated for $J$ and $H$ are
$M_K=\mkbmaxval$ with an RMS of \krms{}~mag.
To check the quality of the light-curve fits
and to see if the fits are robust, we plot
the $H$-band residuals with respect to the \chisqnu{} and 
degrees of freedom of the fits to the $H$-band template in
Fig.~\ref{fig:res_mu_vs_chisq_dof}. 
Fig.~\ref{fig:res_mu_vs_delta_av} 
demonstrates that after an extinction correction of $A_H=0.19 E(B-V)$, 
there is no correlation in the $H$-band
residuals with the light-curve shape parameter, $\Delta$, or extinction, $A_V$,
as measured from the optical light curves by MLCS2k2~\citep{jha07}.

To reduce the effects of uncertainties in the local flow,
we also calculated the RMS 
for only the \numsneiaflow{}~\sneia{} with a CMB+Virgo infall model
velocity of $>2000$~\kms{}.  We found an RMS of \hflowrms{}~mag
for this sample.
Previous work by \citet{radburn-smith04} and \citet{neill07}
found that \snia{} distances were more consistent with the Hubble law when a local flow model
was incorporated.
Future work from larger samples of nearby \sneia{} in the NIR
will allow for the investigation of both the local flow and 
the absolute NIR magnitudes of \sneia{} from integrated analyses
with more sophisticated flow models of the local Universe.

\section{Conclusions}
\label{sec:conc}

We have constructed improved NIR templates to find
accurate luminosity distances to \sneia{} 
using the $H$-band light curve from $[-10,+20]$~days after $B$-band maximum light.  
Within the photometric and local-flow uncertainties, the distribution about a best-fit \lcdm{} model 
is consistent with
no intrinsic dispersion of our $H$-band template fit magnitude values. 
Even without accounting for peculiar velocity uncertainties, the RMS scatter
of \hflowrms{}~mag from these {\em uncorrected} inferred luminosity distance 
moduli is as small as {\em corrected} distances from optical-based methods. 
No correction for light-curve shape has been made here and the treatment
of dust extinction is relatively limited in its extrapolation 
over a factor of 2 in wavelength.  
It is quite plausible that using information on the NIR and optical light-curve shapes, dust absorption as fit from the optical to the infrared, spectroscopic variation, or host-galaxy properties will result in even smaller scatter and better distance determinations.  

The \citet{krisciunas04a} result was determined from an 
inhomogeneous sample of 16 light curves.
This paper improves on that work by providing
a homogeneous sample of \numsneianew{} PAIRITEL \snia{} light curves
that more than doubles the number of independent photometric
observations in the literature.
With our ongoing program at
Mt.~Hopkins adding $\sim10$--$15$ \sneia{} per year, PAIRITEL will
continue building an extensive nearby training set of ground-based NIR \sneia{}.  
While we observe all the \sneia{} that meet our criteria, objects in the Hubble flow
observed with adequate integration times to beat down the measurement errors will be the most helpful.

In this paper, we considered reddening values derived from optical data, 
and found a relationship of apparent $A_H = 0.06 A_V$ 
between the absolute $H$-band magnitude and the optical reddening.
However, recent work on the highly extinguished 
SN~2002cv~\citep{elias-rosa07} shows that reddening corrections
will be necessary for some highly-absorbed \sneia{} in the NIR.  
Reddening uncertainties derived from optical data alone currently
represent the most significant systematic error affecting \snia{}
luminosity distance measurements. Simulations by \citet{krisciunas07}
demonstrate major improvements to be gained from the addition of NIR
data.  \citet{krisciunas07} show that simulated
distance modulus errors are improved by factors of 2.7 and 3.5 by
adding $J$, and \jhk{} to $UBVRI$ data, respectively.
Reducing systematics due to reddening are crucial
to future space-based \snia{} surveys which will be large enough to
avoid limitations from sample size. In that case, improving the constraints 
on cosmological parameters will be limited by systematics. 
We will explore the utility of a full combination of optical and NIR data to more
precisely measure and correct for reddening in \citet{friedman08}.

Although ground-based NIR data can be obtained for low redshift objects,
rest-frame NIR observations for high-redshift \sneia{} will have to be done from space.  
Currently, 
rest-frame \snia{} Hubble diagrams of high-redshift \sneia{} have yet to be
constructed beyond the $I$ band \citep{freedman05,nobili05}, with
limited studies of \sneia{} and their host galaxies conducted in the
mid-infrared with Spitzer \citep{chary05,gerardy07}.

Because nearby \sneia{} are excellent standard candles in the NIR, it may
be worth careful consideration of the rest-frame NIR for the space missions using \sneia{} for cosmology.  
For example, the SNAP~\citep{aldering04} and DESTINY
satellites \citep{lauer05}, candidates for the NASA/DOE
Joint Dark Energy Mission (JDEM) mission, are both currently 
designed with detectors sensitive out to $1.7$~$\mu$m, 
which will only detect rest-frame $H$-band
light ($1.6$~$\mu$m) out to $z \sim 0.1$.  Only a detector capable of
observing rest-frame $H$-band at $z \sim 0.5$--$2$ could take full
advantage of Nature's gift to us of a superb standard candle in the rest frame $H$-band.  
Such a detector would require sensitivity from
$2$--$5$~$\mu$m. 
We can look forward to work of this type with JWST, just over the horizon.
The astronomical community should discuss 
which aspects of a JDEM mission could best be carried out in the rest-frame NIR.

\acknowledgments

The authors thank the anonymous referee for detailed and thorough
reports that substantially improved the clarity of this manuscript.
Thanks to K. Mandel for thorough reading of the paper and catching
several critical typos in the initially submitted draft.
The Peters Automated Infrared Imaging Telescope (PAIRITEL) is operated
by the Smithsonian Astrophysical Observatory (SAO) and was made
possible by a grant from the Harvard University Milton Fund, the
camera loan from the University of Virginia, and the continued support
of the SAO and UC Berkeley. Partial support for PAIRITEL operations
and this work comes from National Aernonautics and Space Administration 
(NASA) grant NNG06GH50G (``PAIRITEL: Infrared Follow-up for Swift Transients''). 
J. S. B. and his group are partially supported by a DOE SciDAC Program through the collaborative agreement DE-FC02-06ER41438.
PAIRITEL support and processing is conducted under the
auspices of a DOE SciDAC grant (DE-FC02-06ER41453), which provides
support to J.S.B.'s group. J.S.B. thanks the Sloan Research Fellowship for
partial support.

We gratefully made use of the NASA/IPAC Extragalactic Database (NED).
This publication makes use of data products from the 2MASS Survey,
funded by NASA and the US National Science Foundation (NSF).
IAUC/CBET were useful. 
M.W.V. is funded by a grant from the US National Science Foundation (AST-057475).
A.S.F. acknowledges support from an NSF Graduate Research Fellowship 
and a NASA Graduate Research Program Fellowship. 
M. M. acknowledges support in part from a Miller Research Fellowship.
A.S.F, R.P.K, M.M., and S.B. thank the Kavli Institute for Theoretical
  Physics, which is supported by the NSF through grant PHY05-51164.
The CfA Supernova Program is supported in part by
NSF Grant AST 06-06772.
C.B. acknowledges support from the Harvard Origins of Life Initiative.

\pagebreak
\bibliographystyle{apj}
\bibliography{sn,pending}

\begin{figure}
\includegraphics[angle=90,width=6in]{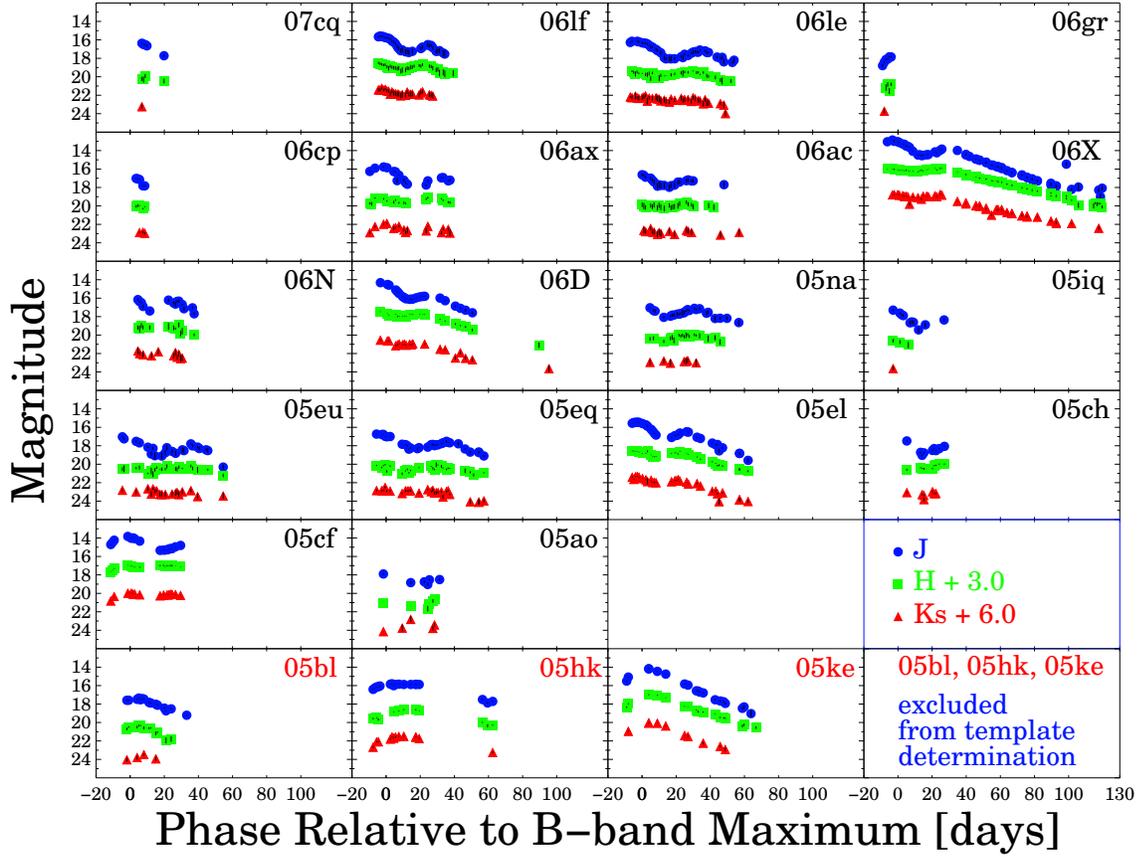}
\caption{PAIRITEL NIR lightcurves of the \numsneianewall{}~\sneia{}  presented in this paper.  There are \numsneianew{} normal \sneia{} plus 3 fast-declining or unusual \sneia{}: SN~2005bl, SN~2005hk, and SN~2005ke.  These \sneia{} are displayed on their own row at the bottom of the plot.}
\label{fig:pairitel_lightcurves}
\end{figure}

\begin{deluxetable}{lrrrrr}
\tablecolumns{6}
\tablewidth{0pc}
\tabletypesize{\scriptsize}
\tablecaption{\snia{} \jhk{} Peak Magnitudes\label{tab:jhkdata}}
\tablehead{
\colhead{SN}
& \colhead{$J_{\rm peak}$\tablenotemark{b}}
& \colhead{$H_{\rm peak}$\tablenotemark{b}}
& \colhead{$K_{\rm peak}$\tablenotemark{b}}
& \colhead{$N_J$,$N_H$,$N_K$\tablenotemark{a}}  & \colhead{Reference\tablenotemark{c}}
}
\startdata
SN 1998bu & 11.55 (0.03) & 11.59 (0.03) & 11.42 (0.03) & 18,20,20 &   J99,H00 \\ 
SN 1999cl & 12.80 (0.02) & 12.77 (0.04) & 12.58 (0.02) &  6, 5, 6 &       K00 \\ 
SN 1999cp & 14.50 (0.02) & 14.77 (0.02) & 14.57 (0.06) &  2, 2, 2 &       K00 \\ 
SN 1999ee & 14.77 (0.03) & 15.00 (0.03) &   \nodata    & 33,33, 0 &      K04a \\ 
SN 1999ek & 16.10 (0.05) & 16.08 (0.04) & 16.54 (0.07) & 14,15, 1 &      K04b \\ 
SN 1999gp & 17.31 (0.06) & 17.16 (0.10) & 16.51 (0.14) &  3, 3, 3 &       K01 \\ 
SN 2000E  & 13.47 (0.02) & 13.74 (0.06) & 13.42 (0.06) & 16,15,10 &       V03 \\ 
SN 2000bh & 16.80 (0.03) & 16.75 (0.03) & 16.64 (0.04) & 20,21,15 &      K04a \\ 
SN 2000bk & 17.64 (0.02) & 17.45 (0.05) &   \nodata    & 18,16, 0 &       K01 \\ 
SN 2000ca & 16.36 (0.03) & 16.71 (0.08) &   \nodata    & 12,11, 0 &      K04a \\ 
SN 2000ce & 16.82 (0.03) & 16.12 (0.03) & 15.98 (0.05) &  3, 5, 5 &       K01 \\ 
SN 2001ba & 16.99 (0.02) & 17.24 (0.04) & 17.06 (0.06) & 15,14, 9 &      K04a \\ 
SN 2001bt & 15.34 (0.07) & 15.64 (0.02) & 15.40 (0.03) & 21,21,11 &      K04b \\ 
SN 2001cn & 15.90 (0.03) & 15.66 (0.05) & 15.67 (0.09) & 19,19, 3 &      K04b \\ 
SN 2001cz & 15.40 (0.06) & 15.65 (0.15) & 15.40 (0.12) & 12,12, 9 &      K04b \\ 
SN 2001el & 12.93 (0.02) & 12.97 (0.03) & 12.84 (0.05) & 33,32,32 &       K03 \\ 
SN 2002bo & 13.68 (0.02) & 13.82 (0.02) & 13.88 (0.04) & 14,14,14 &      K04b \\ 
SN 2002cv & 14.75 (0.03) & 14.34 (0.01) & 13.95 (0.03) & 17,10,16 & DP02,ER07 \\ 
SN 2002dj & 14.56 (0.09) & 14.79 (0.03) & 14.50 (0.09) & 21,21,21 &       P08 \\ 
SN 2003du & 14.42 (0.02) & 14.66 (0.02) & 14.38 (0.02) &  6, 6, 6 &      St07 \\ 
SN 2003cg & 13.58 (0.08) & 13.68 (0.10) & 13.43 (0.01) & 10,10,10 &      ER06 \\ 
SN 2004S  & 14.95 (0.02) & 15.04 (0.02) & 14.87 (0.06) & 16,17,12 &       K07 \\ 
SN 2004eo & 15.70 (0.07) & 15.79 (0.07) & 15.67 (0.09) & 11, 9,10 &     Pa07b \\ 
SN 2005ao & 17.90 (0.08) & 17.65 (0.00) & 16.82 (0.32) &  6, 6, 5 &      WV08 \\ 
SN 2005bl\tablenotemark{d} & 17.35 (0.04) & 17.33 (0.07) & 17.47 (0.14) & 15,12, 4 &  T08,WV08 \\ 
SN 2005cf & 13.83 (0.04) & 13.96 (0.02) & 13.97 (0.02) & 17,17,15 &      WV08\tablenotemark{e} \\ 
SN 2005ch & 17.48 (0.06) & 16.99 (0.05) & 16.20 (0.14) & 10,10, 8 &      WV08 \\ 
SN 2005el & 15.45 (0.02) & 15.55 (0.03) & 15.31 (0.03) & 29,29,29 &      WV08 \\ 
SN 2005eq & 16.74 (0.03) & 17.05 (0.08) & 16.51 (0.11) & 30,26,27 &      WV08 \\ 
SN 2005eu & 17.03 (0.09) & 17.20 (0.16) & 16.64 (0.22) & 22,22,18 &      WV08 \\ 
SN 2005hk\tablenotemark{d} & 15.81 (0.03) & 15.60 (0.05) & 15.48 (0.05) & 18,12,13 &      WV08 \\ 
SN 2005iq & 17.30 (0.09) & 17.61 (0.19) &   \nodata    & 10, 3, 0 &      WV08 \\ 
SN 2005ke\tablenotemark{d} & 14.18 (0.04) & 13.98 (0.03) & 14.03 (0.02) & 19,18,10 &      WV08 \\ 
SN 2005na & 17.04 (0.05) & 16.96 (0.14) & 16.76 (0.27) & 24,19, 7 &      WV08 \\ 
SN 2006D  & 14.32 (0.02) & 14.47 (0.03) & 14.54 (0.05) & 22,22,20 &      WV08 \\ 
SN 2006N  & 16.16 (0.11) & 15.90 (0.26) & 15.72 (0.32) & 14,12,11 &      WV08 \\ 
SN 2006X  & 12.86 (0.03) & 12.92 (0.02) & 12.74 (0.04) & 60,66,47 &      WV08\tablenotemark{e} \\ 
SN 2006ac & 16.62 (0.14) & 16.60 (0.20) & 16.44 (0.18) & 20,22,15 &      WV08 \\ 
SN 2006ax & 15.77 (0.05) & 16.11 (0.10) & 15.89 (0.06) & 17,16,16 &      WV08 \\ 
SN 2006cp & 17.02 (0.04) & 16.96 (0.15) & 16.84 (0.13) &  5, 5, 3 &      WV08 \\ 
SN 2006gr & 17.84 (0.07) & 17.82 (0.18) & 17.72 (0.25) &  6, 4, 1 &      WV08 \\ 
SN 2006le & 16.15 (0.10) & 16.38 (0.05) & 16.05 (0.17) & 38,32,36 &      WV08 \\ 
SN 2006lf & 15.59 (0.08) & 15.50 (0.13) & 15.24 (0.14) & 30,33,26 &      WV08 \\ 
SN 2007cq & 16.37 (0.05) & 16.92 (0.21) &   \nodata    &  5, 4, 0 &      WV08 \\ 
\enddata
\tablenotetext{a} {Number of epochs w/ SNR$>3$ in the \jhk light curves, respectively.}
\tablenotetext{b} {Magnitudes of the maximum observed magnitude in \jhk. 
Magnitude errors are 1-$\sigma$ symmetric errors.}
\tablenotetext{c} {Reference codes: 
WV08 (PAIRITEL photometry; this paper);
J99:  \citet{jha99}; 
H00:  \citet{hernandez00};
K00:  \citet{krisciunas00}; 
DP02: \citet{dipaola02};
V03:  \citet{valentini03};
K03:  \citet{krisciunas03};
K04a: \citet{krisciunas04a}; 
K04b: \citet{krisciunas04b}; 
K07:  \citet{krisciunas07};
ER06: \citet{elias-rosa06};
ER07: \citet{elias-rosa07};
Pa07b: \citet{pastorello07b}; 
St07: \citet{stanishev07};
P08:  \citet{pignata08}.}
\tablenotetext{d}{ The fast-declining, unusual SN~2005bl, SN~2005hk, and SN~2005ke were excluded from the template generation.  When fit with the H-band template they are clearly rejected as outliers with $\chisqnu>3$.  See Fig.~\ref{fig:weirdsneia_vs_template} for a visual comparison of the H-band lightcurves of these dromedary \sneia{} compared with the H-band template.}
\tablenotetext{e}{Because we had data from PAIRITEL on these \sneia{} we did not take advantage of the photometry of SN~2005cf provided in \citet{pastorello07a} or that of SN~2006X provided in \citet{wangx08}.}
\end{deluxetable}

\begin{deluxetable}{rrrrrrr}
\tabletypesize{\scriptsize}
\tablewidth{0pc}
\tablecaption{\snia{} \jhk Templates}
\tablehead{
\colhead{Phase} &
\colhead{$J$ Magnitude} & 
\colhead{Uncertainty} & 
\colhead{$H$ Magnitude} & 
\colhead{Uncertainty} & 
\colhead{$K_s$ Magnitude} & 
\colhead{Uncertainty}  \\
}
\startdata
-10.00 &   0.59 &   0.26 &   0.45 &   0.23 &   0.64 &   0.26 \\
 -9.00 &   0.36 &   0.30 &   0.26 &   0.26 &   0.51 &   0.33 \\
 -8.00 &   0.24 &   0.25 &   0.16 &   0.23 &   0.33 &   0.27 \\
 -7.00 &   0.10 &   0.20 &   0.04 &   0.21 &   0.20 &   0.21 \\
 -6.00 &   0.00 &   0.13 &  -0.03 &   0.18 &   0.10 &   0.09 \\
 -5.00 &  -0.07 &   0.09 &  -0.08 &   0.14 &   0.03 &   0.10 \\
 -4.00 &  -0.11 &   0.07 &  -0.10 &   0.13 &  -0.02 &   0.11 \\
 -3.00 &  -0.11 &   0.07 &  -0.09 &   0.13 &  -0.04 &   0.10 \\
 -2.00 &  -0.09 &   0.09 &  -0.07 &   0.13 &  -0.04 &   0.10 \\
 -1.00 &  -0.06 &   0.11 &  -0.04 &   0.13 &  -0.03 &   0.10 \\
  0.00 &   0.00 &   0.11 &   0.00 &   0.12 &   0.00 &   0.13 \\
  1.00 &   0.07 &   0.13 &   0.04 &   0.11 &   0.04 &   0.14 \\
  2.00 &   0.16 &   0.15 &   0.07 &   0.10 &   0.08 &   0.14 \\
  3.00 &   0.26 &   0.19 &   0.10 &   0.12 &   0.12 &   0.13 \\
  4.00 &   0.37 &   0.24 &   0.13 &   0.13 &   0.15 &   0.13 \\
  5.00 &   0.50 &   0.26 &   0.15 &   0.12 &   0.19 &   0.13 \\
  6.00 &   0.64 &   0.25 &   0.18 &   0.12 &   0.22 &   0.12 \\
  7.00 &   0.78 &   0.26 &   0.20 &   0.14 &   0.25 &   0.14 \\
  8.00 &   0.92 &   0.27 &   0.23 &   0.14 &   0.28 &   0.14 \\
  9.00 &   1.05 &   0.26 &   0.25 &   0.13 &   0.31 &   0.17 \\
 10.00 &   1.17 &   0.26 &   0.26 &   0.13 &   0.33 &   0.18 \\
 11.00 &   1.28 &   0.24 &   0.25 &   0.13 &   0.35 &   0.17 \\
 12.00 &   1.37 &   0.22 &   0.24 &   0.11 &   0.35 &   0.18 \\
 13.00 &   1.43 &   0.19 &   0.22 &   0.12 &   0.34 &   0.17 \\
 14.00 &   1.47 &   0.16 &   0.19 &   0.13 &   0.33 &   0.15 \\
 15.00 &   1.48 &   0.15 &   0.15 &   0.13 &   0.31 &   0.14 \\
 16.00 &   1.47 &   0.16 &   0.12 &   0.11 &   0.30 &   0.17 \\
 17.00 &   1.44 &   0.19 &   0.09 &   0.11 &   0.29 &   0.16 \\
 18.00 &   1.40 &   0.20 &   0.06 &   0.11 &   0.28 &   0.17 \\
 19.00 &   1.36 &   0.23 &   0.03 &   0.13 &   0.26 &   0.18 \\
 20.00 &   1.32 &   0.26 &   0.00 &   0.13 &   0.24 &   0.18 \\
 21.00 &   1.26 &   0.26 &  -0.03 &   0.12 &   0.22 &   0.16 \\
 22.00 &   1.21 &   0.23 &  -0.06 &   0.11 &   0.21 &   0.17 \\
 23.00 &   1.17 &   0.27 &  -0.07 &   0.12 &   0.22 &   0.17 \\
 24.00 &   1.11 &   0.29 &  -0.07 &   0.12 &   0.24 &   0.17 \\
 25.00 &   1.06 &   0.29 &  -0.06 &   0.14 &   0.26 &   0.19 \\
 26.00 &   1.02 &   0.28 &  -0.04 &   0.16 &   0.30 &   0.19 \\
 27.00 &   0.98 &   0.26 &   0.00 &   0.22 &   0.33 &   0.21 \\
 28.00 &   0.96 &   0.26 &   0.03 &   0.24 &   0.35 &   0.21 \\
 29.00 &   0.96 &   0.25 &   0.06 &   0.28 &   0.36 &   0.21 \\
 30.00 &   0.97 &   0.24 &   0.10 &   0.29 &   0.38 &   0.22 \\
 31.00 &   1.00 &   0.25 &   0.13 &   0.25 &   0.40 &   0.25 \\
 32.00 &   1.04 &   0.25 &   0.17 &   0.25 &   0.43 &   0.25 \\
 33.00 &   1.10 &   0.25 &   0.21 &   0.25 &   0.47 &   0.25 \\
 34.00 &   1.16 &   0.25 &   0.26 &   0.25 &   0.52 &   0.25 \\
 35.00 &   1.23 &   0.25 &   0.30 &   0.25 &   0.57 &   0.25 \\
 36.00 &   1.30 &   0.25 &   0.35 &   0.25 &   0.62 &   0.25 \\
 37.00 &   1.37 &   0.25 &   0.40 &   0.25 &   0.67 &   0.25 \\
 38.00 &   1.44 &   0.25 &   0.45 &   0.25 &   0.72 &   0.25 \\
 39.00 &   1.51 &   0.25 &   0.49 &   0.25 &   0.77 &   0.25 \\
 40.00 &   1.58 &   0.25 &   0.54 &   0.25 &   0.82 &   0.25 \\
 50.00 &   2.27 &   0.25 &   1.01 &   0.25 &   1.31 &   0.25 \\
 60.00 &   2.97 &   0.25 &   1.49 &   0.25 &   1.80 &   0.25 \\
 70.00 &   3.67 &   0.25 &   1.96 &   0.25 &   2.30 &   0.25 \\
 80.00 &   4.37 &   0.25 &   2.44 &   0.25 &   2.79 &   0.25 \\
\enddata

\tablecomments{The templates are linear fits from $+35$~days to $+80$~days with ($J$, $H$, $K_s$) slopes of $(0.0625, 0.0470, 0.425)$~mag/day.}
\label{tab:jhk_templates}
\end{deluxetable}

\begin{deluxetable}{lrrlllllrrllll}
\rotate
\tabletypesize{\scriptsize}
\tablewidth{0pc}
\tablecaption{Apparent Standard $H$-Band Magnitudes of Type Ia Supernovae}
\tablehead{
\colhead{SN Name} 
  & \colhead{$cz_{\rm CMB+Virgo}$\tablenotemark{a}} & \colhead{$cz_{\rm err}$} & \colhead{$\sigma_\mu({\rm vel})$\tablenotemark{b}} 
  & \colhead{\tbmax} & \colhead{$H_{\rm B max}$\tablenotemark{c}} & \colhead{$\sigma(H_{\rm B max})$} 
  & \colhead{\chisqnu} & \colhead{DoF\tablenotemark{d}} & \colhead{$A_V$} & \colhead{$\mu$ \tablenotemark{e}} & \colhead{$\sigma_\mu$\tablenotemark{f}} \\
\colhead{} 
  & \colhead{\kms}                                  & \colhead{\kms}               & \colhead{[mag]} 
  & \colhead{[MJD]}  & \colhead{[mag]}  & \colhead{[mag]}              & \colhead{[mag]} 
  & \colhead{}                     & \colhead{}         & \colhead{[mag]}                   & \colhead{} \\
}
\startdata
SN~1998bu   &             709 &       20 &   0.463 &    50952.4 &   11.76 &    0.04 &    0.42 &   12 &  1.12 &   29.76 &    0.46  \\
SN~1999cl   &             957 &       86 &   0.392 &    51342.2 &   12.76 &    0.06 &    1.42 &    3 &  3.49 &   30.77 &    0.40  \\
SN~1999cp   &            2909 &       14 &   0.112 &    51363.2 &   14.89 &    0.11 &    0.20 &    1 &  0.08 &   32.90 &    0.16  \\
SN~1999ee   &            3296 &       15 &   0.099 &    51469.3 &   15.11 &    0.03 &    0.14 &   13 &  0.96 &   33.12 &    0.11  \\
SN~1999ek   &            5191 &       10 &   0.063 &    51481.8 &   16.35 &    0.04 &    0.30 &   12 &  0.54 &   34.36 &    0.07  \\
SN~1999gp   &            8113 &       18 &   0.040 &    51550.1 &   17.07 &    0.10 &    0.24 &    2 &  0.10 &   35.08 &    0.11  \\
SN~2000E    &            1803 &       19 &   0.182 &    51577.2 &   13.84 &    0.05 &    0.51 &    8 &  0.66 &   31.84 &    0.19  \\
SN~2000bh   &            6765 &       21 &   0.049 &    51636.0 &   16.85 &    0.04 &    0.78 &    7 &  0.16 &   34.86 &    0.06  \\
SN~2000bk   &            7976 &       20 &   0.041 &    51647.7 &   17.35 &    0.03 &    0.50 &   12 &  0.30 &   35.36 &    0.05  \\
SN~2000ca   &            6989 &       62 &   0.050 &    51666.2 &   16.78 &    0.04 &    0.74 &    9 &  0.03 &   34.79 &    0.06  \\
SN~2000ce   &            5097 &       15 &   0.064 &    51667.1 &   16.13 &    0.07 &    0.36 &    2 &  1.87 &   34.14 &    0.09  \\
SN~2001ba   &            8718 &       22 &   0.038 &    52034.2 &   17.52 &    0.05 &    0.51 &    8 &  0.05 &   35.53 &    0.06  \\
SN~2001bt   &            4220 &       13 &   0.077 &    52062.9 &   15.84 &    0.05 &    0.42 &    9 &  0.69 &   33.84 &    0.09  \\
SN~2001cn   &            4454 &      250 &   0.142 &    52071.0 &   15.82 &    0.05 &    0.25 &    6 &  0.52 &   33.83 &    0.15  \\
SN~2001cz   &            4506 &       20 &   0.073 &    52103.4 &   15.84 &    0.07 &    0.05 &    6 &  0.30 &   33.85 &    0.10  \\
SN~2001el   &             978 &       10 &   0.334 &    52182.1 &   13.00 &    0.05 &    0.38 &    8 &  0.79 &   31.01 &    0.34  \\
SN~2002bo   &            1696 &       20 &   0.194 &    52356.0 &   13.92 &    0.05 &    0.15 &    6 &  1.53 &   31.93 &    0.20  \\
SN~2002cv   &            1696 &       20 &   0.194 &    52417.7 &   13.87 &    0.06 &    0.43 &    7 &  8.74 &   31.88 &    0.20  \\
SN~2002dj   &            2880 &       22 &   0.114 &    52450.6 &   14.86 &    0.04 &    0.20 &   10 &  0.41 &   32.87 &    0.12  \\
SN~2003cg   &            1340 &       24 &   0.246 &    52729.1 &   13.56 &    0.07 &    0.03 &    4 &  4.20 &   31.57 &    0.25  \\
SN~2003du   &            2206 &       14 &   0.148 &    52766.2 &   14.62 &    0.05 &    0.60 &    4 &  0.04 &   32.63 &    0.16  \\
SN~2004S    &            2607 &       16 &   0.126 &    53038.7 &   14.95 &    0.05 &    0.24 &    5 &  0.40 &   32.96 &    0.13  \\
SN~2004eo   &            4859 &       17 &   0.067 &    53278.7 &   15.91 &    0.07 &    0.30 &    4 &  0.01 &   33.92 &    0.10  \\
SN~2005ao   &           11828 &      126 &   0.036 &    53442.0 &   18.10 &    0.09 &    0.08 &    1 &  0.10 &   36.11 &    0.09  \\
SN~2005bl   &            7360 &       29 &   0.045 &    53482.1 &   17.44 &    0.05 &    3.65 &    9 &  0.22 &   35.45 &    0.07  \\
SN~2005cf   &            2018 &       11 &   0.162 &    53533.6 &   14.00 &    0.04 &    0.14 &    7 &  0.29 &   32.01 &    0.17  \\
SN~2005ch   &            8094 &     1499 &   0.404 &    53536.0 &   17.26 &    0.06 &    0.73 &    5 & \nodata &   35.27 &    0.41  \\
SN~2005el   &            4349 &        8 &   0.075 &    53646.1 &   15.79 &    0.03 &    0.78 &   14 &  0.03 &   33.80 &    0.08  \\
SN~2005eq   &            8535 &       25 &   0.039 &    53653.9 &   17.34 &    0.04 &    2.11 &   11 &  0.11 &   35.35 &    0.06  \\
SN~2005eu   &           10503 &       14 &   0.031 &    53659.8 &   17.38 &    0.05 &    0.87 &   11 &  0.08 &   35.39 &    0.06  \\
SN~2005hk   &            3855 &       21 &   0.085 &    53685.7 &   15.68 &    0.04 &   13.09 &    8 &  1.19 &   33.69 &    0.10  \\
SN~2005iq   &           10102 &       40 &   0.033 &    53687.1 &   17.73 &    0.15 &    0.06 &    2 &  0.05 &   35.74 &    0.15  \\
SN~2005ke   &            1230 &       11 &   0.266 &    53699.3 &   14.05 &    0.06 &    8.32 &    5 &  0.34 &   32.06 &    0.27  \\
SN~2005na   &            7826 &       26 &   0.042 &    53740.5 &   17.23 &    0.10 &    0.28 &    5 &  0.08 &   35.24 &    0.11  \\
SN~2006D    &            2560 &       18 &   0.128 &    53756.7 &   14.67 &    0.03 &    0.33 &   12 &  0.11 &   32.68 &    0.13  \\
SN~2006N    &            4468 &       27 &   0.074 &    53760.6 &   15.94 &    0.12 &    0.23 &    4 &  0.12 &   33.95 &    0.15  \\
SN~2006X    &            1091 &       20 &   0.301 &    53785.5 &   12.79 &    0.03 &    0.24 &   17 &  3.83 &   30.80 &    0.30  \\
SN~2006ac   &            7123 &       17 &   0.046 &    53781.2 &   16.88 &    0.08 &    0.23 &   13 &  0.08 &   34.89 &    0.09  \\
SN~2006ax   &            4955 &       20 &   0.066 &    53826.7 &   16.36 &    0.05 &    0.38 &    9 &  0.14 &   34.37 &    0.09  \\
SN~2006cp   &            6816 &       14 &   0.048 &    53896.7 &   16.83 &    0.07 &    0.63 &    4 &  0.75 &   34.84 &    0.09  \\
SN~2006gr   &           10547 &       22 &   0.031 &    54012.2 &   18.12 &    0.15 &    1.16 &    3 &  0.19 &   36.12 &    0.15  \\
SN~2006le   &            5403 &       12 &   0.060 &    54047.2 &   16.64 &    0.05 &    0.38 &   16 &  0.09 &   34.65 &    0.08  \\
SN~2006lf   &            4048 &       10 &   0.081 &    54044.8 &   15.81 &    0.05 &    0.18 &   21 &  0.26 &   33.82 &    0.09  \\
SN~2007cq   &            7501 &       50 &   0.046 &    54272.5 &   16.97 &    0.14 &    1.26 &    3 & \nodata &   34.98 &    0.15  \\
\enddata

\tablenotetext{a} {Redshift of host galaxies as corrected to CMB frame
with the additional correction of Virgo member NGC~4501 (host of
SN~1999cl) from its observed CMB-corrected recession velocity (2281
km/s) to the mean recession velocity of Virgo (957 km/s).}

\tablenotetext{b} {A peculiar velocity of 150~km/s and the individual
redshift measurement uncertainty of column 2 was converted into an
equivalent distance modulus uncertainty using Eq.~\ref{eq:dztodmu}.}

\tablenotetext{c} {An H-band extinction correction of $A_H=0.06 A_V$  has been applied to the magnitudes obtained by fitting to the template.  See \S\ref{sec:standard} for details.}

\tablenotetext{d} {DoF: The number of degrees of freedom is the number
of $H$-band data points minus 1 for the overall offset fit parameter,
\hbmax.}

\tablenotetext{e} {Assuming $H_0=72$~\kmsmpc and thus $M_H=\mhbmaxval$~mag.}

\tablenotetext{f} {Quadrature sum of peculiar velocity distance modulus uncertainty (column 4) and fit uncertainty (column 7).}

\label{tab:hmax}
\end{deluxetable}

\begin{deluxetable}{lrrrr}
\tablewidth{0pc}
\tablecaption{JHK Absolute Magnitudes of \sneia{} at \tbmax}
\tablehead{
\colhead{Filter} & \multicolumn{2}{c}{PAIRITEL sample} & \multicolumn{2}{c}{Global sample} \\
\colhead{}       & \colhead{M} & \colhead{RMS} & \colhead{M} & \colhead{RMS} 
}
\startdata
J & \mjbmaxptelval & \jptelrms & \mjbmaxval & \jrms \\ 
H & \mhbmaxptelval & \hptelrms & \mhbmaxval & \hrms \\
K & \mkbmaxptelval & \kptelrms & \mkbmaxval & \krms \\
\enddata
\tablecomments{The J-band template shows significant variation as a function of optical light-curve width.  The numbers derived above are for the average template as shown in Fig.~\ref{fig:template}.  The K-band data from PAIRITEL is only useful for 9 new \sneia{} and does not significantly improve on the results of \citet{krisciunas04a}.}

\end{deluxetable}

\begin{deluxetable}{lllrr}
\tablewidth{0pc}
\tablecaption{Light Curve Table Stub}
\tablehead{
\colhead{SN} & \colhead{MJD} & \colhead{Passband} & \colhead{Flux$_{25}$\tablenotemark{a}} & \colhead{Flux err$_{25}$} \\
}
\startdata
2005ao   &  53440.4157        &        H   &  600.2491  &     3.5737 \\
2005ao   &  53456.4300        &        H   &  437.9836  &     3.6405 \\
2005ao   &  53464.3872        &        H   &  339.2386  &   128.0434 \\
2005ao   &  53466.3879        &        H   &  345.1047  &    90.6351 \\
2005ao   &  53467.3847        &        H   &  561.7854  &     3.7594 \\
\multicolumn{4}{l}{\nodata } \\
\enddata
\tablenotetext{a} {Fluxes are expressed normalized to a zeropoint of 25.  I.e., the calibrated 2MASS magnitude is mag$=-2.5\log_{10}({\rm flux})+25$}
\tablecomments{This is a representative stub of the light-curve table.  \jhk{} light curves for all \numsneianewall{} PAIRITEL \sneia{} presented in this paper are available in the online electronic version.}
\label{tab:irlcs}
\end{deluxetable}

\end{document}